\documentclass[iop,revtex4]{emulateapj}
\usepackage{amsmath}
\usepackage{footnote}
\usepackage{textcomp}
\usepackage[T1]{fontenc}
\usepackage{mathtools}
\usepackage{rotating}
\usepackage[flushleft]{threeparttable}
\usepackage{natbib}
\usepackage{lscape}
\usepackage[percent]{overpic}

\bibpunct{(}{)}{;}{a}{}{,}
\usepackage{hyperref}
\hypersetup{
    colorlinks,
    citecolor=blue,
    filecolor=blue,
    linkcolor=blue,
    urlcolor=blue
}
\slugcomment{Accepted by ApJ May 2018} 
\shortauthors{Allen et al.}

\begin{document}

\title{The Disk Wind in the Neutron Star Low-mass X-Ray Binary GX 13+1}
\author{Jessamyn L. Allen\altaffilmark{1,2}}
\author{Norbert S. Schulz\altaffilmark{2}}
\author{Jeroen Homan\altaffilmark{2,3}}
\author{Joseph Neilsen\altaffilmark{2}}
\author{Michael A. Nowak\altaffilmark{2}}
\author{Deepto Chakrabarty\altaffilmark{1,2}}

\altaffiltext{1}{Department of Physics, Massachusetts Institute of Technology, 77 Massachusetts Avenue, Cambridge, MA 02139, USA; allenjl@alum.mit.edu}
\altaffiltext{2}{Kavli Institute for Astrophysics \& Space Research, MIT, 70 Vassar St., Cambridge, MA 02139, USA}
\altaffiltext{3}{SRON, Netherlands Institute for Space Research, Sorbonnelaan 2, 3584 CA Utrecht, The Netherlands}

\begin{abstract}
We present the analysis of seven \emph{Chandra} High Energy Transmission Grating Spectrometer and six simultaneous \emph{RXTE} Proportional Counter Array observations of the persistent neutron star (NS) low-mass X-ray binary GX 13+1 on its normal and horizontal branches. Across nearly 10 years, GX 13+1 is consistently found to be accreting at $50-70$\% Eddington, and all observations exhibit multiple narrow, blueshifted absorption features, the signature of a disk wind, despite the association of normal and horizontal branches with jet activity. A single absorber with standard abundances cannot account for all seven major disk wind features, indicating that multiple absorption zones may be present. Two or three absorbers can produce all of the absorption features at their observed broadened widths and reveal that multiple kinematic components produce the accretion disk wind signature. Assuming the most ionized absorber reflects the physical conditions closest to the NS, we estimate a wind launching radius of $7\times10^{10}$ cm, for an electron density of $10^{12}$ cm$^{-3}$. This is consistent with the Compton radius and also with a thermally driven wind. Because of the source's high Eddington fraction, radiation pressure likely facilitates the wind launching.
\end{abstract}

\section{Introduction}
\label{sec:intro}
Warm absorbers (partially ionized gas) are common to both black hole (BH) and neutron star (NS) low-mass X-ray binaries (LMXBs), identified through the narrow absorption features imprinted on the illuminating continuum. A group of warm absorbers have only been seen in sources with high inclinations, indicating the absorbing material is associated with the accretion disk. In all BH LMXBs and a subset of the NS binaries, the absorption features are blueshifted \citep{pontibh12,trigoboirin12,trigoboirin,ponti16}; the absorbing material forms an outflow known as an accretion disk wind. When blueshifts are not present, the absorber is in a static configuration and is commonly called an accretion disk atmosphere. Warm absorbers were first observed in active galactic nuclei (AGNs) in the 1970s and later in X-ray binaries with the launch of \emph{ASCA} in the 1990s \citep{ueda98,kotani2000}. The high-resolution gratings onboard \emph{Chandra} and \emph{XMM} subsequently revealed the complexity and ubiquity of the warm absorbers in LMXBs and, in the case of accretion disk winds, showed them to be a major component of the accretion picture \citep{neilsen11,pontibh12,ponti14}.

The warm absorbers' ionized resonance features are most often transitions in highly ionized species of iron, including the K$\alpha$ line of \ion{Fe}{25} and \ion{Fe}{26}, although some sources' spectra exhibit over 90 absorption lines from transitions in dozens of ionized species, such as the BH binary GRO J1655$-$40 \citep{miller06,kallman09}. A single absorption zone can often explain the absorption features, but in a number of sources, especially those with features other than iron lines, multiple absorption zones can more accurately reproduce the complex absorption signature. In the case of accretion disk winds, outflow speeds are typically in the range $300-1000$ km s$^{-1}$, although ultrafast outflows with velocities of $0.01-0.04c$ have been reported \citep{miller15multabs,miller16gx340,miller16}. As the absorption features are only seen in high-inclination LMXBs ($60^{\circ}<i< 80^{\circ}$ constrained by the presence of dips in light curves; \citealt{frank87}) and no significant line re-emission is observed, the absorbers are believed to have an equatorial geometry with a small opening angle or possibly a bipolar geometry with strong stratification in density and/or ionization above the accretion disk plane \citep{trigoboirin12,higproga15}.

In BH LMXBs there is a strong correlation between mass outflow and the accretion state. Accretion disk winds detected via blueshifted absorption features are found almost exclusively in soft spectral states, when the accretion disk is optically thick and the spectrum is disk-dominated \citep{pontibh12}. Jets, detected via strong radio emission, are active in the hard spectral states, when the inner disk is believed to be optically thin and the spectrum has significant nonthermal emission. The blueshifted, narrow absorption features are also absent in hard states of high-inclination sources, suggesting the accretion disk wind is absent. There are, however, significant exceptions to the outflow$-$accretion state correlation; there is strong evidence that in at least five disk wind sources, including BH LMXB GRS 1915+105, winds and jets are simultaneous in luminous hard spectral states \citep{lee01,homan16}.

NSs may show a similar correlation between accretion state and outflow \citep{ponti14,bianchi17}, but only 30\% of high-inclination NS warm absorbers exhibit definite outflows \citep{trigoboirin}. In addition, NS LMXB outburst tracks are more varied than those of BHs \citep{nsstates}, and several NSs are included in the subset of disk wind systems with possible simultaneous jets and winds \citep{homan16}. NS LMXBs are classified by their behavior as atoll or Z sources, the latter being brighter ($L_{\mathrm{X}}>0.5$ $L_{\mathrm{Edd}}$) and less variable. In general, NSs are softer than BHs in outburst, in part due to the thermal boundary layer emission, but even NSs show significant variety in spectral softness depending on their classification, with Z sources being much softer overall than atoll sources. These issues compound to make the outflow$-$spectral state correlation less obvious in NS LMXBs compared to BHs.

Radiation pressure, Compton heating, and magnetic forces can all launch outflows in accretion disk systems. In NS and BH LMXBs, the warm absorber plasmas are highly ionized, leaving too few transitions in the UV and the soft X-ray for line driving to be an effective wind launching mechanism \citep{progakallman02}. In a Compton-heated outflow,\footnote{While Compton-heated winds are often called \textquotedbl thermally driven winds\textquotedbl\, we will use \textquotedbl Compton-heated winds\textquotedbl, or simply \textquotedbl Compton winds\textquotedbl \ to be explicit about the heating mechanism.}, the accretion disk is strongly irradiated by the central X-ray-emitting region; beyond some point in the disk, the thermal velocity exceeds the local escape velocity, generating an outflow. Several BH LMXBs, including GRO J1655-40 and GRS 1915+105 \citep{miller06,miller16}, have exhibited fast and/or dense outflows. Magnetorotational instabilities or magnetocentrifugal forces can launch winds very close to the central compact object, which can account for the high wind densities and large blueshifts. There is also the possibility that different wind driving mechanisms are active at the same time but dominate in different spectral states and/or luminosities \citep{neilsen12,homan16}, due to a hybrid wind driving mechanism. For the brightest disk wind sources accreting at near-Eddington rates, radiation pressure on electrons should increasingly contribute to the wind driving in addition to the dominant mechanism \citep{progakallman02,homan16,done16,ryota18}.

Determining disk wind launching mechanisms and spectral state dependence have powerful implications beyond understanding disk wind properties. While disk winds are relatively slow outflows ($v\lesssim0.01c$), they can carry significant mass away from the binary, comparable to the accretion rate. High mass-loss rates can create accretion disk instabilities which may produce luminosity modulations or instabilities that may drive state transitions \citep{begel83,shields86}, and may also be significant enough to alter binary and spin evolution timescales \citep{pontibh12}. A complete picture of outflows in stellar mass compact object systems will help tp understand accretion on much larger mass scales, including supermassive BHs in AGNs which also exhibit powerful jets and winds.

\subsection{NS LMXB GX 13+1}
\label{sec:introsrc} 

The NS LMXB GX 13+1 (also known as 4U 1811$-$171) is a bright, persistent accretor located in the Galactic bulge. It orbits an evolved late-type K5 III star \citep{fleisch85, bandy99} and has an estimated distance of $7\pm1$ kpc \citep{bandy99}. Studies of infrared and X-ray light curve modulations \citep{corbet10,iaria14} as well as X-ray dips \citep{dai14dip}, support a binary orbital period of 24.7 days and a highly inclined orbit ($60^{\circ}-80^{\circ}$; \citealt{dtrigoxmm}).

Originally classified as an atoll source \citep{hasinger89}, GX 13+1 was later re-classified as a Z source due to strong secular evolution of its color$-$color and hardness$-$intensity diagrams (CDs and HIDs), rapid movement along its CD \& HID tracks, and variability levels, all behaviors consistent with other well-established Z sources \citep{homan98,frid15gx13z}. 

A warm absorber in GX 13+1 was first discovered by \citet{ueda2001}, made evident by an iron K$\alpha$ absorption line in GX 13+1's \emph{ASCA} spectrum. Subsequent \emph{Chandra} and \emph{XMM} observations revealed a more detailed picture of the ionized material's properties via its absorption imprinted on GX 13+1's spectrum. Analyzing a set of \emph{XMM-Newton} European Photon Imaging Camera (EPIC) observations, \citet{sidoli2002} reported K$\alpha$ and K$\beta$ absorption lines of \ion{Fe}{25} and \ion{Fe}{26}, along with a K$\alpha$ \ion{Ca}{20} line. Despite large velocity shifts detected in several absorption features ($-$3000 to $-$5000 km s$^{-1}$), blueshifts were not detected in all absorption lines and the errors on the outflow velocity were large ($\pm$ 2200 km s$^{-1}$). A \emph{Chandra} High Energy Transmission Grating Spectrometer (HETGS) observation in 2001 of GX 13+1 revealed the K$\alpha$ lines from H-like Fe, Mn, Cr, Ca, Ar, S, Si, and Mg as well as He-like Fe \citep{ueda2004}. The absorption lines shared a common blueshift of $\approx460$ km s$^{-1}$, corresponding to an outflow velocity of $\approx400$ km s$^{-1}$ when corrected for proper motion. The inferred mass outflow rate of the disk wind was determined to be comparable to the accretion rate ($\approx10^{18}$ g s$^{-1}$), indicating disk winds are a significant component of the accretion process in GX 13+1.

\citet{ueda2004} favored a radiation-driven outflow that develops at sub-Eddington luminosities for the accretion disk wind in GX 13+1. In later \emph{XMM} RGS observations of GX 13+1 \citet{dtrigoxmm} found strong correlations between the hard flux and the warm absorber's properties (column density and ionization), concluding the absorber was consistent with a Compton-heated disk wind.

\citet{madej14} analyzed all available \emph{Chandra} gratings observations of GX 13+1 in an attempt to measure orbital parameters using the disk wind, but were hindered by the intrinsic variability of the source and the wind. In this paper we present a re-analysis of the same set of archived \emph{Chandra} HETGS observations with the primary goal of understanding the evolution of this wind. We perform detailed absorption line fitting and utilize \emph{RXTE} data to track the disk wind properties throughout the system's CD. In contrast to previous studies that assumed a single absorber, we find that the disk wind must comprise multiple absorption zones.

\section{Observations}
\label{sec:observ}
\begin{table*}
   \caption{ \textbf{\emph{Chandra} HETGS and \emph{RXTE} PCA Observation Details} \label{tab:cxoobsummary} }
   \tabcolsep=0.125cm
   \tabletypesize{\footnotesize}
   \centering
\begin{threeparttable}
\begin{tabular}{l | c c c c | c c c}
   \hline
   \hline
 Date &  & \multicolumn{2}{c}{\emph{Chandra} HETGS} & & & \multicolumn{1}{c}{\emph{RXTE} PCA}  & \\
 & ObsID & Mode & Exposure (ks) & Count Rate (s$^{-1}$)\tnote{a} & Exposure (ks) & Count Rate (s$^{-1}$)  & \emph{RXTE} Color Diagram Position\tnote{b} \\
   \hline
2002 Oct 8 &2708   &  TE & $29.4$ & $13.5$ & $9.8$ & $16.7$ & Upper NB $\rightarrow$ NB/HB vertex \\
2010 Jul 24 & 11815 &  TE & $28.1$ & $13.9 $ & $13.2$ & $16.7 $ & Mid NB $\rightarrow$ Upper NB \\
2010 Jul 30 & 11816 &  TE & $28.1$ & $13.6 $ & $5.7$ & $16.1$ & Lower HB $\rightarrow$ Upper NB \\
2010 Aug 1 & 11814 &  TE & $26.8/28.1$\tnote{c} & $11.2/11.1$ & $9.1$\tnote{d} & $17.0$ & NB/HB vertex$\rightarrow$ upper HB  \\
2010 Aug 3 & 11817 &  TE & $28.1$ & $13.0$ & $10.3$ & $16.4$ & NB/HB vertex $\rightarrow$ NB/FB vertex \\
2010 Aug 5 & 11818 &  CC & $23.0$ & $14.6$ & $5.2$ & $15.8$ & Lower NB $\rightarrow$ NB/HB vertex \\
2011 Feb 17 & 13197 &  CC & 10.1 & 17.9 & $\cdots$ & $\cdots$& $\cdots$ \\
   \hline
\end{tabular}
\begin{tablenotes}
    \item[a] The \emph{Chandra} $0.5-10$ keV count rate and the \emph{RXTE} $3-40$ keV count rate.
    \item[b] NB=Normal Branch, HB=Horizontal Branch, FB=Flaring Branch
    \item[c] Observation details when  filtered/unfiltered for the dip event.
    \item[d] The \emph{RXTE} observation did not cover the dipping event and, hence, was not filtered.
\end{tablenotes}
\end{threeparttable}
\end{table*}

In our study of GX 13+1's accretion disk wind, we utilize a dataset of seven \emph{Chandra} HETGS observations, six of which have simultaneous \emph{RXTE}/Proportional Counter Array (PCA) coverage. The observations' details are listed in Table \ref{tab:cxoobsummary}.

\subsection{\emph{Chandra} Observations}

We have analyzed all seven \emph{Chandra} HETGS \citep{canizares} observations of GX 13+1, for a total exposure of approximately 190 ks. The observations span 10 years, with five of the observations taken during a two week timespan in 2010 July and August. Observations were taken in both timed exposure (TE) and continuous clocking (CC) modes with the Advance CCD Imaging Spectrometer S-array (ACIS-S). For TE mode observations, a 350-row subarray was used, yielding a 1.24 s frame time. In CC mode, spatial information was collapsed into 1 row, reducing the frame time to 2.85 ms. 

The \emph{Chandra} data were downloaded from the \emph{Chandra} Data Archive\footnote{http://cxc.harvard.edu/cda/} and re-processed using the TGCat \citep{tgcat} \emph{run\_pipe} script with CIAO 4.9 \citep{ciao} and CALDB 4.5.0 in \texttt{ISIS}\footnote{http://space.mit.edu/cxc/isis/} version 1.6.2-30 \citep{isis}. Our GX 13+1 HETGS data processed with CIAO 4.9 differ slightly from data processed with older versions of CIAO due to changes in the correction for contaminant build-up on \emph{Chandra's} optical blocking filter. The most significant deviations approach the 10\% level at long wavelengths ($>8$ \AA) for both the High Energy Grating (HEG) and Medium Energy Grating (MEG) first orders.

The zero-order position determines the absolute accuracy of the wavelength measurements. As GX 13+1's zeroth order is significantly piled-up in TE mode observations, we cannot use a centroid calculation due to the distorted point spread function. We used \emph{findzo} to calculate the zero-order position from the intersection of the MEG spectrum and the ACIS frame-shift streak, which provides a zero-order position accuracy of less than 0.1 ACIS-S detector pixels, corresponding to 0.001 \AA \ for the HEG and MEG first orders. First-order redistribution matrix files (RMFs) and ancillary response files (ARFs) were created with the \emph{run\_pipe} script.

\citet{dai14dip} reported a dipping event approximately halfway through ObsID 11814. Dips have been associated with spectral changes in both the warm absorber and neutral absorption column \citep{diaztrigowarm}. We removed the dip event, including the ingress and egress of the event based on times reported by \citet{dai14dip} and performed spectral analysis only on the non-dipping spectrum. 

We used the AGLC script\footnote{http://space.mit.edu/cxc/analysis/aglc/aglc.html} in \texttt{ISIS} to create the ACIS-S grating light curves and compute hardness ratios. Hardness ratios are defined as the number of hard counts ($3-8$ keV) divided by the number of soft counts ($0.5-3$ keV). The count rate varied on the level of 15\%, but we found no significant changes in the source's intensity or spectral hardness, so we analyzed observations whole, except for ObsID 11814 as stated above.

\subsection{\emph{RXTE} Observations}

We also analyzed data from the \emph{RXTE} PCA \citep{rxtepca}. For a list of the individual \emph{RXTE}/PCA observations that were performed simultaneously with our \emph{Chandra} observations we refer to Table 1 in \citet{homan16}. As the \emph{RXTE} continuum parameters did not vary significantly between orbits, we summed the individual orbit event files using the \emph{sumpha} tool to produce a single \emph{RXTE} spectrum for each corresponding \emph{Chandra} HETGS observation.

\begin{figure}
   \centering
\begin{tabular}{c}
   \includegraphics[scale=0.95,angle=0]{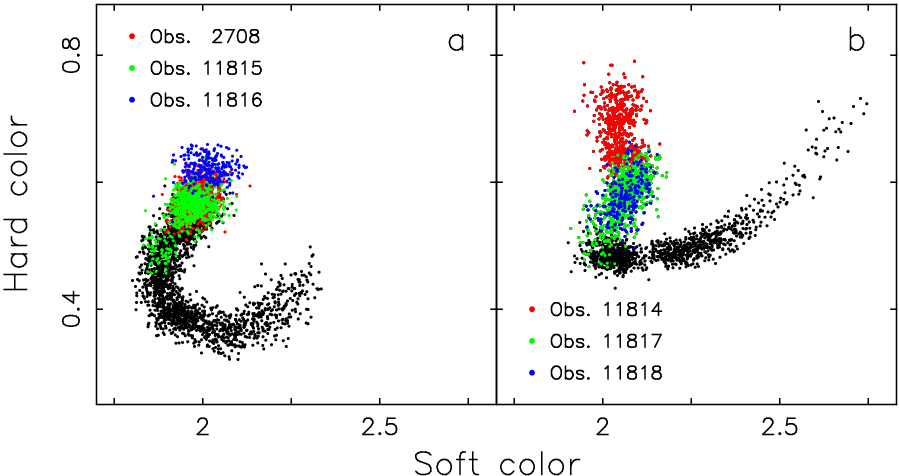} \\
\end{tabular}
   \caption{Two of GX 13+1's Z track CDs with the color-coded points corresponding to six of the \emph{Chandra} observations that occurred simultaneously with \emph{RXTE} monitoring. In several observations, including ObsID 11816, the source only moved slightly along the Z track during the course of the observation (from the lower HB to the upper NB), while in ObsID 11817 the source moved along the entire horizontal branch, from the NB/HB vertex to the NB/FB vertex. \protect{\label{fig:ztrack}} }
\end{figure}

For our X-ray CDs of GX 13+1, we made use of \emph{RXTE}/PCA data published in \citet{frid15gx13z}. We refer to that paper for the details of the data reduction. GX 13+1 exhibits significant secular motion of its Z track on the timescales of days or longer. \citet{frid15gx13z} organized the source's \emph{RXTE} data into six different CD tracks. Correlating the source's \emph{RXTE} position on the Z tracks during our \emph{Chandra} HETGS observations, see Figure \ref{fig:ztrack}, ObsIDs 2708, 11815, and 11816 fall on one Z track that has less distinct branches \citep{homan16}, while GX 13+1 moved along the horizontal branch (HB) and normal branch (NB) on another Z track during ObsIDs 11814, 11817, and 11818.

\section{Analysis and Results}
\label{sec:analysis} 
All spectral analysis was performed within \texttt{ISIS}. Errors on fit parameters were calculated with \emph{conf\_loop} and correspond to the 90\% confidence bounds ($\sigma=1.6$, $\Delta \chi^2 = 2.71$). All quoted chi-squared values, $\chi^2_{\nu} \ (dof)$, are reduced. We used the \emph{calc\_flux} \texttt{ISIS} tool to calculate (absorbed) fluxes and (unabsorbed) luminosities. The $0.5-10$ keV and bolometric ($0.1-100$ keV) luminosities were calculated assuming a distance of 7 kpc. The Eddington fraction ($L_{\mathrm{X}} / L_{\mathrm{Edd}}=f_{\mathrm{Edd}}$) assumed an Eddington luminosity, $L_{\mathrm{Edd}}$, of $2 \times 10^{38}$ erg s$^{-1}$ for a 1.4 $M_{\odot}$ NS with a hydrogen-rich photosphere.

We fit the uncombined HEG and MEG $\pm$1 orders with matched spectral grids across the $1.65-9.5$ \AA \ ($1.3-7.5$ keV) range, corresponding to a resolution of 0.023\ \AA\ across our energy band. Each bin contained a minimum of 20 counts which allowed us to use $\chi^2$ statistics.

While the HEG has twice the spectral resolution as the MEG and experiences less pile-up, our spectral analysis benefited from the additional MEG counts. Continuum and line parameters were better constrained when both HEG and MEG first orders ($\pm$ 1) were used\footnote{The only exceptions are the \ion{Fe}{26} and\ion{Fe}{25} absorption lines and the broad iron emission line, for which we excluded the MEG due to its low effective area at 7 keV.}. ObsID 13197 was a CC mode calibration observation; two arms (MEG $-$1 and HEG $+$1) fell off the CCD, leaving only the MEG $+$1 and HEG $-$1 orders.

For the \emph{RXTE}/PCA data, the background subtracted spectra were extracted from the Proportional Counter Unit 2 and a systematic error of 0.6\% was added. Bins between $3$ and $40$ keV were noticed in the continuum fits.

\subsection{Continuum Fitting}
\label{sec:contfit}

\subsubsection{Pile-up}
\label{sec:pileup}

As GX 13+1 is bright ($\approx$0.2 Crab) and five of the seven observations were taken in TE mode, spectra can suffer from significant pile-up. Pile-up has effects of energy and event grade migration that reduce the total source count rate and distort the observed spectral shape\footnote{http://cxc.harvard.edu/ciao/download/doc/pileup\_abc.pdf}. We used the \texttt{ISIS} convolution model \emph{simple\_gpile2} to quantify and mitigate the pile-up effects \citep{nowaksimple,hankesimple}. The TE mode observations experienced maximum pile-up at 3 \AA\ ($\approx 4.1$ keV).

In addition to generating spectral distortion, pile-up has the effect of \emph{reducing} an absorption feature's measured equivalent width (EW). The local pile-up fraction is proportional to the local count rate, which means that the local continuum is more suppressed relative to the absorption feature, reducing the observed EW. Our attempt to correct the EWs due to pile-up suppression is outlined in Section \ref{sec:gaussfit}.

While the CC mode observations have a significantly shorter frame time making pile-up essentially negligible, they do not help us quantify the pile-up effects on the continua of the TE mode observations due to contamination by an X-ray scattering halo. GX 13+1 is highly absorbed and the CC mode continua include the dispersed scattering halo, which is difficult to model.

\subsubsection{ISM Absorption}

The neutral absorption column toward GX 13+1 is known to be large, with published values in the range of $N_{\mathrm{H}}=(3-5)\times10^{22}$ cm$^{-2}$ ($N_{\mathrm{H},22}=3-5$; \citealt{ueda2001,dtrigoxmm,schulzsik}). We used the absorption model \emph{tbnew} v2.3, with the abundances set to those of \citet{wilmsabund} and the cross-sections set to \citet{verner}.

In our joint fits of the \emph{Chandra} HETGS and \emph{RXTE}/PCA data, we found that the $\chi^2_{\nu}$ was improved when we allowed the column density to vary between the two datasets. The \emph{Chandra} HETGS data were well-fit with column densities in the range of $N_{\mathrm{H},22}=4.9-5.0$, while for the \emph{RXTE}/PCA data, $N_{\mathrm{H},22}=3.5-3.9$. The \emph{Chandra} HETGS column density was consistently $\approx30\%$ higher than that of the \emph{RXTE}/PCA column density and is likely associated with the X-ray scattering halo. \emph{Chandra's} field of view is narrow relative to \emph{RXTE}'s and, thus, the \emph{Chandra} HETGS observations do not include the X-ray scattering halo, while the \emph{RXTE}/PCA observations do. The scattered X-rays from the halo are an additional source of low-energy flux in the \emph{RXTE} spectrum, which can be modeled, to first order, with a lower neutral column density. We leave a more complete investigation of the X-ray scattering halo's effects on the joint analysis of \emph{Chandra} and \emph{RXTE} data to a future paper.

We also found that the $\chi^2_{\nu}$ value was always improved when the silicon abundance of the HETGS \emph{tbnew} component was allowed to vary, yielding silicon overabundances between 2.0 and 2.1. The required silicon abundance is due to \emph{tbnew}'s incomplete modeling of the Si K edge \citep{schulzsik}. Additional structure in the Si K-edge, i.e. the near and far edge absorption, is apparent, including a line at 1.865 keV likely associated with a moderately ionized plasma in GX 13+1. We did not attempt to model the silicon edge beyond allowing for the overabundance in \emph{tbnew}.

\subsubsection{Continuum Models}
\label{sec:contmod}

\newpage

\begin{sidewaystable}
   \centering
   \caption{\textbf{Parameters for DiskBB+BB+PL Continuum Model} \label{tab:contparam} }
   \tabcolsep=0.0875cm
   \scriptsize
\begin{threeparttable}
\begin{tabular}{l c c c c c c c c c c c c c c r}
   \hline
   \hline
ObsID & $N_{\mathrm{H},22}(1)$\tnote{a} & Si (1)\tnote{b} & $N_{\mathrm{H},22}(2)\tnote{c} $ & BB kT & BB norm\tnote{d} & Disk T$_{\mathrm{in}}$ & Disk Norm\tnote{d} & PL  norm\tnote{e} & $\beta$ & $\beta$  & $F_{0.5-10}$ (BB) \tnote{f}  & $F_{8-10}$\tnote{g} &  $L_{0.5-10}$\tnote{h}  &  $f_{\mathrm{Edd}}$\tnote{i}  & $\chi_{\nu}^2$ (dof) \\
 & & & & (keV) & (km$^{2}$)  & (keV) & (km$^{2}$) &  & $\pm$1 MEG & $\pm$1 HEG  &   &  &  &  & \\
   \hline
2708 & $4.88\pm0.04$ & $2.1\pm0.1$ & $3.87\pm0.27$ & $2.61\pm0.09$ & $1.5\pm0.3$ & $1.75\pm0.02$ & $94\pm4$ & $0.31\pm0.10$ & $0.031/0.031$ & $0.017/0.020$ & $6.93 \ (0.10)$ & $0.90$ & $8.18$ & $0.45$ & $1.156 \ (6047)$  \\ 
11815 & $4.97\pm0.03$ & $2.1\pm0.1$ & $3.59\pm0.25$ & $2.36\pm0.09$ & $2.2\pm0.6$ & $1.69\pm0.02$ & $105\pm5$ & $0.51\pm0.06$ & $0.025/0.026$ & $0.017/0.017$ & $7.02 \ (0.11)$ & $0.87$ & $8.55$ & $0.45$ & $1.137 \ (5975)$  \\ 
11816 & $4.91\pm0.04$ & $2.1\pm0.1$ & $3.50\pm0.26$ & $2.48\pm0.08$ & $2.6\pm0.6$ & $1.72\pm0.03$ & $95\pm5$ & $0.46\pm0.10$ & $0.026/0.028$ & $0.019/0.019$ & $7.07 \ (0.15)$ & $0.96$ & $8.35$ & $0.45$ & $1.141 \ (5939)$  \\ 
11814 & $4.99\pm0.03$ & $2.0\pm0.1$ & $3.48\pm0.26$ & $2.45\pm0.07$ & $3.1\pm0.5$ & $1.70\pm0.03$ & $76\pm4$ & $0.62\pm0.09$ & $0.029/0.030$ & $0.024/0.023$ & $6.02 \ (0.20)$ & $0.88$ & $7.25$ & $0.38$ & $1.123 \ (5831)$  \\ 
11817 & $5.00\pm0.03$ & $2.0\pm0.1$ & $3.55\pm0.25$ & $2.57\pm0.08$ & $1.4\pm0.3$ & $1.82\pm0.02$ & $77\pm1$ & $0.36\pm0.08$ & $0.027/0.028$ & $0.019/0.019$ & $6.81 \ (0.09)$ & $0.91$ & $7.93$ & $0.43$ & $1.134 \ (5883)$  \\
   \hline
\end{tabular}
   \begin{tablenotes}
   \item[a] \emph{Chandra} HETGS column density.
   \item[b] Allowing for a silicon overabundance in \emph{Chandra} HETGS \emph{tbnew} component improved the Si K edge fit.
   \item[c] \emph{RXTE}/PCA column density.
   \item[d] The emission areas of the thermal components were calculated assuming a distance of $d=7$ kpc. For the disk component, we additionally assumed an inclination of $i=70^{\circ}$.
   \item[e] The powerlaw normalization. The powerlaw photon index, $\Gamma$, is fixed to 2.5 in all fits.
   \item[f] The $0.5-10$ keV \emph{Chandra} HETGS flux in units of $10^{-9}$ erg s$^{-1}$ cm$^{-2}$ with the blackbody fraction in parentheses.
   \item[g] The $8-10$ keV \emph{Chandra} HETGS flux in units of $10^{-9}$ erg s$^{-1}$ cm$^{-2}$.
   \item[h] The $0.5-10$ keV \emph{Chandra} HETGS luminosity with units $10^{37}$ erg s$^{-1}$ (assuming $d=7$ kpc).
   \item[i] The Eddington fraction calculated from the $0.1-100$ keV \emph{RXTE}/PCA unabsorbed flux; the powerlaw flux is excluded below the blackbody temperature.
   \end{tablenotes}
\end{threeparttable}
\end{sidewaystable}

\begin{figure*}[t!]
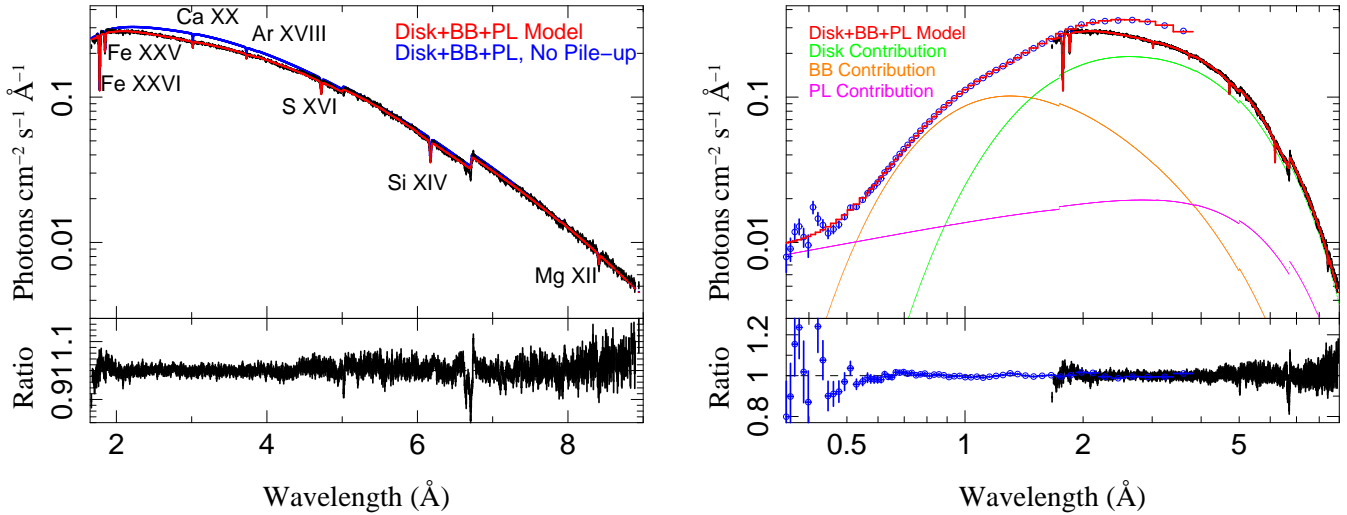

   \centering
\begin{tabular}{c c}
   \includegraphics[angle=270,scale=0.35]{te_combined_tbnew_diskbb_lines_em_paper.ps} &
   \includegraphics[angle=270,scale=0.35]{te_rxte_combined_tbnew_diskbb_lines_em_paper.ps} \\
\end{tabular}
   \caption{Left: combined HEG and MEG $\pm$ 1 orders for all five TE mode observations (data plotted in black) fit with a disk plus blackbody continuum and gaussian absorption features (shown in red). The continuum model without \emph{simple\_gpile2}, plotted in blue, shows how the spectrum would appear without pile-up. Additional structure in the silicon edge is seen in the residuals near 6.6 \AA. Right: The continuum fit to the combined TE mode observations (black) along with the combined \emph{RXTE}/PCA observations (blue). The disk, blackbody, and powerlaw contributions are plotted in green, orange, and purple, respectively.  \protect{\label{fig:diskbbcontpic}}}
\end{figure*}

When fit by itself, the GX 13+1 \emph{Chandra} HETGS continuum was consistent with multiple model prescriptions as the large interstellar medium (ISM) absorption ($N_{\mathrm{H},22} >1$) reduces the soft X-ray sensitivity and limits our ability to distinguish between continuum models. We found the \emph{diskbb+bbodyrad} and \emph{bbodyrad+powerlaw} continuum models, the most common models for NS LMXBs accreting at high rates, performed equally well.

Joint fits of the HETGS and PCA spectrum required three major spectral components; we fit a \emph{diskbb+bbodyrad+powerlaw} model to the broadband continuum. Both the HETGS and PCA spectra were modified by neutral absorption columns with variable column densities. The HETGS continuum also included the pile-up convolution model \emph{simple\_gpile2} and seven narrow absorption features modeled with negative Gaussians (see Section \ref{sec:gaussfit}).

The PCA spectrum was modified by two edge components: one with a fixed energy ($8.83$ keV) and another left free to vary, yielding an average fit value of $7.1$ keV. Both edges were required to obtain good $\chi^2_{\nu}$ values for our fits of the \emph{RXTE} data. The edge at 8.83 keV is unresolved in the PCA spectrum but is seen in the \emph{XMM} GX 13+1 spectrum \citep{dtrigoxmm}, while the edge at 7.1 keV is likely a superposition of narrow iron absorption features and a broad iron emission line. Details of fitting a broad emission line to the \emph{Chandra} data are in Section \ref{sec:broadem}.

The continuum fit to the TE mode observations is shown in Figure \ref{fig:diskbbcontpic} and the continuum parameters can be found in Table \ref{tab:contparam}. In our continuum prescription, the multicolor disk component is the accretion disk while the blackbody component is likely boundary layer emission from the NS. Both the blackbody and disk components exhibit slight variations in their temperatures and normalizations, but there is no clear correlation among the parameters or the source flux, likely due to the large neutral column which introduces degeneracies in our continuum fits. The $0.5-10$ keV absorption-corrected luminosity varies on the order of 10\% across all observations, $L_{\mathrm{X}} \approx(7.3-8.6)\times10^{37}$ erg s$^{-1}$.

\subsubsection{Broad Iron Emission Line}
\label{sec:broadem}

Broad iron emission lines are common to both BH and NS LMXBs \citep{white86,asai2000}. In NS binaries, possible origins include the inner accretion disk, an accretion disk corona, and an ionized accretion disk wind, with the line being broadened by relativistic, Compton scattering and electron down-scattering mechanisms, respectively.

A broad iron emission line has previously been seen in the spectra of GX 13+1 with multiple X-ray instruments including \emph{ASCA}, \emph{XMM}/EPIC, \emph{RXTE}/PCA and \emph{Chandra} HETGS \citep{asai2000,ueda2001,sidoli2002,ueda2004,dtrigoxmm,dai14dip}. Although nearby Fe K$\alpha$ absorption features make it difficult to constrain the emission line's parameters, its energy and EW have been found to be variable. A 1994 \emph{ASCA} observation revealed an emission line with energy $6.42\pm0.08$ keV, $\sigma<$ 220 eV, and  EW=19$\pm$8 eV \citep{asai2000,ueda2001}, while in \emph{XMM} observations the emission line had higher energies ($6.5-6.8$ keV) and significantly larger widths ($\sigma=0.7-0.9$ keV) and EWs ($100-200$ eV) \citep{sidoli2002,dtrigoxmm}. The correlation between the iron emission and absorption line EWs suggested a common origin in the outer disk (i.e. an accretion disk wind; \citealt{dtrigoxmm}).

\begin{figure}
   \centering
\begin{tabular}{c}
   \includegraphics[scale=0.6, angle=270]{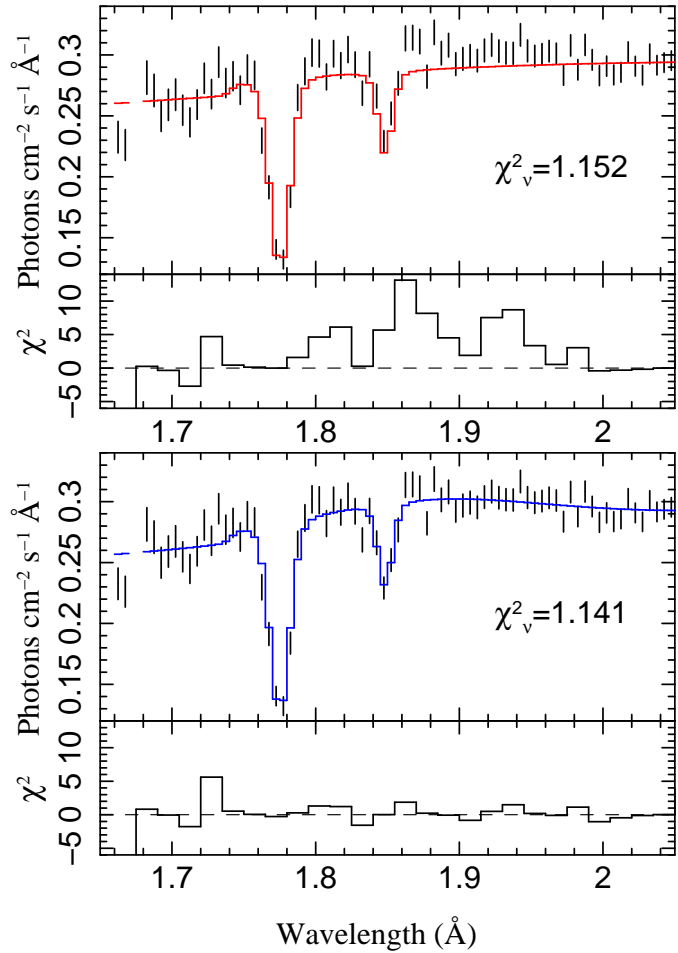} \\
\end{tabular}
   \caption{ObsID 11816, MEG and HEG $\pm$1 orders. Top: when only the local \ion{Fe}{26} and \ion{Fe}{25} absorption features are fit, there is an excess of positive residuals in the $1.8-2.0$ \AA\ range. The residuals have been heavily binned to make the residual pattern more apparent. Bottom: adding an emission line centered at 1.88 \AA\ (6.60 keV, indicated by the dashed vertical line) improves the fit. \protect{\label{fig:feresid}} }
\end{figure}

The \emph{Chandra} HETGS data are well fit in the $6-7$ keV ($\approx1.8-2$ \AA ) range with two narrow absorption features corresponding to the K$\alpha$ transitions of \ion{Fe}{25} and \ion{Fe}{26} (See Section \ref{sec:gaussfit}), but an excess of positive residuals is apparent (see the top panel of Figure \ref{fig:feresid}). The \emph{RXTE}/PCA data cannot resolve the narrow iron absorption features but require an "edge-like" feature at 7.1 keV; this feature does not appear as an edge in the \emph{Chandra} or \emph{XMM} data.

The best joint fit to the \emph{Chandra} and \emph{RXTE} data is achieved when a broad emission line is added to the \emph{Chandra} continuum model, along with the two narrow absorption features, while only an edge is fit to the \emph{RXTE} data. The improved fit to the \emph{Chandra} data with the emission line is shown in the bottom panel of Figure \ref{fig:feresid}.

We allowed the \emph{RXTE}/PCA edge energy to vary even though fixing the edge energy did not affect the quality of the fits. Simply fitting two narrow absorption features and a broad emission line to both the \emph{Chandra} and \emph{RXTE} did not work, implying there is additional absorption outside our \emph{Chandra} band that is unresolved with \emph{RXTE}. \citet{dtrigoxmm} reported additional narrow iron absorption features at $7.91$ and $8.24$ keV.
 
The broad emission line's properties are listed in Table \ref{tab:febroadparam}. We found the broad line had typically energies of $6.48-6.60$ keV ($1.88-1.91$ \AA), $\sigma=0.18-0.29$ keV, and EWs of $20-40$ eV, more consistent with the emission line properties when observed with \emph{ASCA} \citep{asai2000,ueda2001} than with \emph{XMM}/EPIC even considering the large \emph{XMM} error bounds \citep{dtrigoxmm}. We see a weak correlation between the broad emission feature's EW and the source's hard flux which has also been seen in \emph{XMM}/EPIC data of GX 13+1; the broad emission line's EW is smallest (largest) in 11814 (11816), when the source is at its faintest (brightest) luminosity in this set of observations. We do not, however, observe the correlation between the narrow iron absorption features' EWs and the broad emission line's EW reported by \citet{dtrigoxmm}.

\begin{table}
   \caption{ \textbf{Broad Fe emission gauss parameters} \label{tab:febroadparam} }
   \centering
   \tabcolsep=0.1cm
   \tabletypesize{\footnotesize}
\begin{tabular}{l c c c r}
   \hline
   \hline
ObsID & $E_C$ & $\sigma$ & Norm  & EW  \\
 &  (keV) & (keV) & ($10^{-3}$ ph s$^{-1}$ cm$^{-2}$)  & (eV) \\
   \hline
2708 & $6.58\pm0.13$ & $0.29\pm0.15$ & $2.27\pm1.12$ & $29\pm14$ \\
11815 & $6.54\pm0.08$ & $0.21\pm0.07$ & $2.26\pm0.80$ & $28\pm10$ \\ 
11816 & $6.60\pm0.07$ & $0.25\pm0.08$ & $3.05\pm0.98$ & $38\pm12$ \\ 
11814 & $6.53\pm0.09$ & $0.18\pm0.10$ & $1.49\pm0.68$ & $21\pm10$ \\ 
11817 & $6.48\pm0.10$ & $0.26\pm0.11$ & $2.30\pm0.86$ & $28\pm11$ \\
   \hline
\end{tabular}
\end{table}

\subsection{Narrow Line Fits}
\label{sec:gaussfit}

\begin{table}
   \caption{ \textbf{Gauss line parameters} \label{tab:gfittab} }
   \centering
   \tabcolsep=0.11cm
   \tabletypesize{\footnotesize}
\begin{tabular}{l l r r r }
   \hline
   \hline
ObsID & Line & $v_{\mathrm{out}}$  &  $v_{\mathrm{turb}}$ & EW \\
	& 	&  (km s$^{-1}$) &  (km s$^{-1}$) & (eV)   \\
   \hline
2708 & Fe-XXVI & $610 \pm 80$ & $1210\pm150$ & $48.0\pm5.0$ \\ 
 & Fe-XXV & $860 \pm 160$ & $1250\pm280$ & $24.0\pm5.0$ \\ 
 & Ca-XX & $550 \pm 140$ & $410\pm210$ & $2.9\pm0.8$ \\ 
 & Ar-XVIII & $440 \pm 180$ & $210\pm100$ & $1.2\pm0.6$ \\ 
 & S-XVI & $480 \pm 70$ & $450\pm140$ & $2.9\pm0.6$ \\ 
 & Si-XIV & $490 \pm 50$ & $460\pm90$ & $2.6\pm0.3$ \\ 
 & Mg-XII & $310 \pm 160$ & $600\pm260$ & $1.6\pm0.8$ \\ 
11815 & Fe-XXVI & $740 \pm 80$ & $1050\pm150$ & $44.0\pm4.0$ \\ 
 & Fe-XXV & $980 \pm 140$ & $1220\pm240$ & $26.0\pm4.0$ \\ 
 & Ca-XX & $550 \pm 160$ & $100\pm50$ & $2.5\pm0.7$ \\ 
 & Ar-XVIII & $620 \pm 200$ & $300\pm150$ & $1.3\pm0.7$ \\ 
 & S-XVI & $610 \pm 110$ & $590\pm230$ & $2.6\pm0.6$ \\ 
 & Si-XIV & $590 \pm 60$ & $540\pm110$ & $2.6\pm0.4$ \\ 
 & Mg-XII & $560 \pm 120$ & $370\pm180$ & $1.1\pm0.7$ \\ 
11816 & Fe-XXVI & $930 \pm 100$ & $1200\pm170$ & $41.0\pm4.0$ \\ 
 & Fe-XXV & $1210 \pm 70$ & $50\pm20$ & $11.0\pm3.0$ \\ 
 & Ca-XX & $450 \pm 750$ & $800\pm400$ & $1.5\pm0.9$ \\ 
 & Ar-XVIII & $530 \pm 280$ & $370\pm180$ & $1.0\pm0.7$ \\ 
 & S-XVI & $880 \pm 230$ & $30\pm10$ & $1.1\pm0.6$ \\ 
 & Si-XIV & $670 \pm 80$ & $230\pm120$ & $1.3\pm0.3$ \\ 
 & Mg-XII & $80 \pm 900$ & $810\pm410$ & $0.7\pm3.0$ \\ 
11814 & Fe-XXVI & $730 \pm 130$ & $1140\pm240$ & $34.0\pm5.0$ \\ 
 & Fe-XXV & $930 \pm 220$ & $320\pm160$ & $11.0\pm4.0$ \\ 
 & Ca-XX & $830 \pm 370$ & $390\pm190$ & $1.5\pm0.8$ \\ 
 & S-XVI & $390 \pm 160$ & $440\pm220$ & $1.7\pm0.7$ \\ 
 & Si-XIV & $270 \pm 80$ & $530\pm140$ & $2.2\pm0.5$ \\ 
 & Mg-XII & $280 \pm 290$ & $540\pm270$ & $1.0\pm1.0$ \\ 
11817 & Fe-XXVI & $500 \pm 90$ & $810\pm180$ & $34.0\pm4.0$ \\ 
 & Fe-XXV & $590 \pm 230$ & $1240\pm380$ & $17.0\pm5.0$ \\ 
 & Ca-XX & $560 \pm 290$ & $200\pm100$ & $1.4\pm0.7$ \\ 
 & Ar-XVIII & $280 \pm 360$ & $450\pm220$ & $1.2\pm0.7$ \\ 
 & S-XVI & $390 \pm 100$ & $110\pm50$ & $1.9\pm0.5$ \\ 
 & Si-XIV & $380 \pm 50$ & $330\pm100$ & $2.2\pm0.3$ \\ 
 & Mg-XII & $320 \pm 100$ & $50\pm20$ & $0.8\pm0.5$ \\ 
11818 & Fe-XXVI & $390 \pm 100$ & $530\pm270$ & $31.0\pm3.0$ \\ 
 & Fe-XXV & $810 \pm 210$ & $700\pm350$ & $15.0\pm3.0$ \\ 
 & Ca-XX & $430 \pm 310$ & $840\pm420$ & $2.9\pm1.0$ \\ 
 & Ar-XVIII & $690 \pm 380$ & $110\pm50$ & $0.6\pm0.3$ \\ 
 & S-XVI & $370 \pm 150$ & $210\pm100$ & $1.6\pm0.6$ \\ 
 & Si-XIV & $340 \pm 80$ & $230\pm120$ & $1.4\pm0.4$ \\ 
 & Mg-XII & $330 \pm 210$ & $210\pm110$ & $0.6\pm0.3$ \\ 
13197 & Fe-XXVI & $870 \pm 200$ & $1210\pm460$ & $40.0\pm7.0$ \\ 
 & Fe-XXV & $880 \pm 220$ & $1280\pm440$ & $33.0\pm6.0$ \\ 
 & Ca-XX & $630 \pm 280$ & $620\pm310$ & $4.0\pm2.0$ \\ 
 & Ar-XVIII & $370 \pm 250$ & $640\pm320$ & $2.8\pm1.0$ \\ 
 & S-XVI & $450 \pm 160$ & $470\pm240$ & $3.1\pm1.0$ \\ 
 & Si-XIV & $560 \pm 90$ & $570\pm140$ & $3.5\pm0.6$ \\ 
 & Mg-XII & $240 \pm 320$ & $570\pm290$ & $2.0\pm1.0$ \\ 
   \hline
\end{tabular}
\end{table}

We focused on known absorption features in GX 13+1's spectrum, i.e. those identified by \citet{ueda2004}. We combined all HEG and MEG $m=\pm 1$ orders using the \texttt{ISIS} \emph{combine\_datasets} function. Using a powerlaw to fit the local continuum within $\pm$ 1 \AA\ of the line center, we fit a gaussian function to the absorption feature ignoring nearby absorption features and edges. Line shifts, turbulent velocities (computed from the gaussian FWHM) and EWs were calculated for each absorption feature in each observation.

\begin{figure}
   \centering
\begin{tabular}{c}
   \includegraphics[angle=270,scale=1.0]{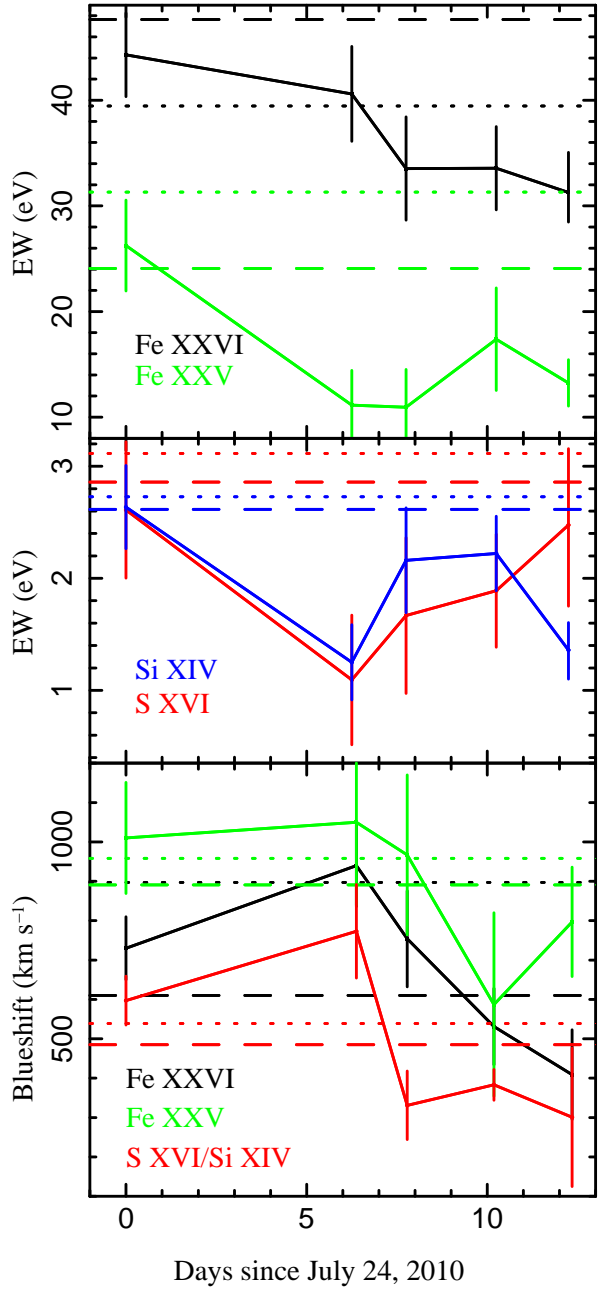} \\
   \caption{Top and middle: The equivalent widths (EWs) of the \ion{Fe}{26} (black), \ion{Fe}{25} (green), \ion{S}{16} (red), and \ion{Si}{14} (blue) absorption features in the observations spanning two weeks in 2010 July$-$August (ObsIDs 11815, 11816, 11814, 11817, and 11818). The long-dashed lines show each absorption line's EW in the 2001 observation (ObsID 2708); the dotted lines indicate values measured in the 2011 observation (ObsID 13197). Bottom: line blueshifts calculated from gaussian fits to the \ion{Fe}{26} (black) and \ion{Fe}{25} (green) absorption lines and an average blueshift of the \ion{S}{16} and \ion{Si}{14} lines (red). There is evidence of an offset between the faster iron lines and the slower silicon and sulfur lines, as well as an offset between the \ion{Fe}{26} and \ion{Fe}{25} lines. \protect{\label{fig:gaussvar}} }
\end{tabular}
\end{figure}

As previously mentioned in Section \ref{sec:pileup}, pile-up reduces the measured EW. To estimate the pile-up suppression, we calculated each absorption line's EW with and without the \emph{simple\_gpile2} pile-up component evaluated in our broadband continuum model. The ratio of the the EWs without and with the pile-up model was a multiplicative correction factor to the EW calculated with the method outlined in the previous paragraph, which provided the best local continuum fit and, hence, the best measure of the EW. As visible in Figure \ref{fig:diskbbcontpic}, pile-up is most severe near 3 \AA, agreeing with our correction factors which were largest for the \ion{Ca}{20} and \ion{Ar}{18} EWs (approximately 10\% and 5\%, respectively) at 3.01 and 3.73 \AA. For the other absorption features, the EWs were suppressed by less than 5\%. The EWs in Table \ref{tab:gfittab} and Figure \ref{fig:gaussvar} have been corrected for pile-up.

Our gaussian line parameters are shown in Table \ref{tab:gfittab}. The K$\alpha$ transitions of \ion{Fe}{26}, \ion{Fe}{25}, \ion{Ca}{20}, \ion{S}{16}, \ion{Si}{14} and \ion{Mg}{12} were detected in all of the \emph{Chandra} HETGS observations. The K$\alpha$ \ion{Ar}{18} is absent in ObsID 11814. The abundance of absorption features in the GX 13+1 spectra is in contrast to the \emph{XMM} observations analyzed in \citet{dtrigoxmm}, where the only disk wind absorption features present were the the K$\alpha$ and K$\beta$ transitions of \ion{Fe}{26} and \ion{Fe}{25}. The \emph{Chandra} HETGS observations lack the signal-to-noise ratio (S/N) above 8 keV to study the iron K$\beta$ transitions, if present. 

The \ion{Fe}{26}, \ion{Fe}{25}, \ion{S}{16} and \ion{Si}{14} EWs across the two week observation period in 2010 July and August are shown in the top and middle panels of Figure \ref{fig:gaussvar}. The solid dashed lines indicate the line's EW in the 2001 observation (ObsID 2708) and the dotted line shows the line's EW in 2011 (ObsID 13197). Despite the observations occurring almost eight years apart, ObsID 2708 (2002 October) and ObsID 11815 (2010 July) exhibit all four major absorption features with almost identical EWs.

There is no clear timescale of variability, but there are significant changes in the absorption features' properties between observations, which occur, at minimum, approximately two days apart. Between ObsIDs 11815 and 11816, separated by six days, the \ion{Fe}{25}, \ion{S}{16}, and \ion{Si}{14} EWs all decrease by factors of $\approx2$ while the \ion{Fe}{26} EW is relatively constant. If all of the absorption features are produced by a single absorption zone, this trend in the EWs requires two simultaneous changes in the accretion disk wind: an increase in the plasma's ionization state along with a decrease in the column density; this would allow the \ion{Fe}{26} line EW to remain constant while producing a decrease in the \ion{S}{16}, \ion{Si}{14}, and \ion{Fe}{25} EWs. Alternatively, if the \ion{Fe}{26} absorption feature is produced by a different absorber than the lower ionization absorption features, a decrease in the column density of the respective absorbing region could produce the observed decrease in the \ion{S}{16}, \ion{Si}{14}, and \ion{Fe}{25} EWs while not necessarily producing any significant change in the \ion{Fe}{26} EW.

We compared the blueshifts of the \ion{Fe}{25} and \ion{Fe}{26} lines with an average of the \ion{S}{16} and \ion{Si}{14} line shifts, shown in the bottom panel of Figure \ref{fig:gaussvar}. The iron lines appear to exhibit larger blueshifts ($500-1200$ km s$^{-1}$) and turbulent velocities (1000 km s$^{-1}$) than absorption features with lower ionization energies, including \ion{S}{16} and \ion{Si}{14} (with $v_{\mathrm{out}}\approx300-600$ km s$^{-1}$ and $v_{\mathrm{turb}}\approx300-550$ km s$^{-1}$). This trend may be related to the decreasing spectral resolution at lower wavelengths in the iron region or may indicate distinct absorption zones with different outflow velocities in the disk wind. Additionally, the \ion{Fe}{25} line exhibits a consistently higher blueshift than the \ion{Fe}{26} line, which could indicate that multiple kinematic components produce the iron absorption features. However, the error bounds on the velocities prevent us from claiming the existence of two or more kinematic components. We do see systematic changes in the outflowing plasma's blueshift between observations, as previously reported by \citet{madej14}. The average blueshift between the two sets of lines increases from $\approx675$ to $\approx850$ km s$^{-1}$ between July 24 and 30 (ObsIDs 11815 and 11816) while the average blueshift decreases to $\approx350$ km s$^{-1}$ six days later (ObsID 11818).

\subsubsection{Manganese and Chromium Lines}

\begin{figure}
   \centering
\begin{tabular}{c}
   \includegraphics[scale=0.3, angle=270.]{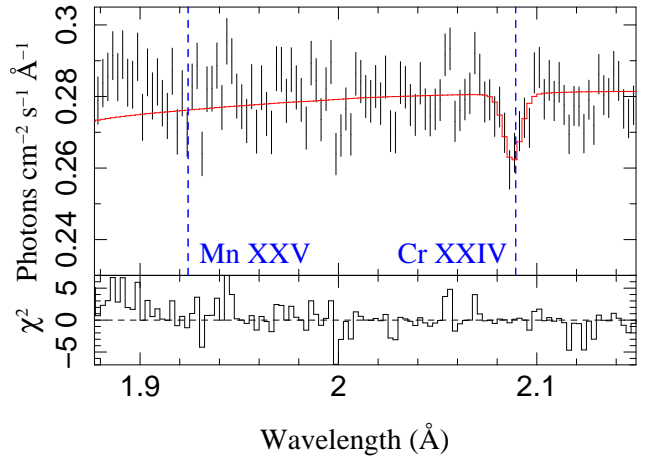} \\
\end{tabular}
   \caption{Combined HEG $\pm$ 1 orders for the five TE mode observations. Rest wavelengths of the \ion{Mn}{25} and \ion{Cr}{24} K$\alpha$ transitions are indicated by the dashed vertical lines. No \ion{Mn}{25} absorption is observed, while a chromium absorption feature can be found at a blueshift of approximately $700$ km s$^{-1}$.\protect{\label{fig:mncrplot}} }
\end{figure}

In the 2001 \emph{Chandra} observation of GX 13+1 (ObsID 2708), \citet{ueda2004} reported \ion{Mn}{25} and \ion{Cr}{24} absorption with EWs of 3.6 and 4.1 eV, respectively. In our analysis of ObsID 2708, we were able to fit absorption features with EWs of 3.5$\pm0.9$ and $3.4^{+0.9}_{-1.6}$ eV but we consider the manganese line consistent with noise. As ObsID 2708 exhibits similar continuum parameters and fluxes as the other observations, we combined all TE mode HEG $\pm1$ orders to increase the S/N in the $1.9-2.1$ \AA\ band, see Figure \ref{fig:mncrplot}. No \ion{Mn}{25} absorption is present, while a possible narrow ($\sigma< 0.005$ \AA) \ion{Cr}{24} line is observed at 2.088 \AA; the line energy corresponds to a blueshift of $\approx 700$ km s$^{-1}$, which is consistent with the blueshifts we observed in the individual TE mode observations ($500-1000$ km s$^{-1}$).

\subsection{Photoionization Model}
\label{sec:warmabsfit}

In essentially all LMXB disk winds, the absorbers have temperatures below the threshold of a collisionally ionized plasma ($k_B \ T_e \approx E_I$, where $T_e$ is the electron temperature, $E_I$ is the ionization energy of the plasma ions). In our \emph{Chandra} HETGS spectra, we modeled the absorption line features with the photoionized plasma \emph{warmabs} (v.2.27) multiplicative component in \texttt{ISIS}. Simultaneously fitting the continuum and the absorption features of the \emph{Chandra} data, the combined function followed the prescription: \emph{simple\_gpile2 [ tbnew $\times$ (warmabs $\times$ (diskbb+bbodyrad+powerlaw)) + emission ]}. The \emph{RXTE} data were included in our fits, but the \emph{warmabs} model was not applied to the PCA spectrum.

\subsubsection{XSTAR Parameters}
\label{sec:inputspec}

The \emph{warmabs} model fits the plasma's bulk properties: ionization state, column density, outflow velocity, and turbulent velocity broadening. The ionization state is characterized by the ionization parameter: 

\begin{equation}\label{eqn:ionparam}
   \xi=\frac{L}{n_e R^2},
\end{equation}

\noindent which depends on the ionizing luminosity ($L$), the electron density ($n_e$), and the distance from the ionizing source ($R$) \citep{tarter}. 

Ion level populations used in our \emph{warmabs} fits were calculated with XSTAR v2.33 \citep{baukall01,kallbau01}. XSTAR simulates a spherical gas shell with a uniform density, illuminated by a central ionizing source. We ran XSTAR with a column density of $N_{\mathrm{H}}=10^{17}$ cm$^{-2}$ to remain within the optically thin limit ($\lesssim10^{24}$ cm$^{-2}$ for ionization parameters $2.5 \leq \log \xi \leq 5$).

Measurements of the disk wind plasma density are rare, as the vast majority of the absorption features are not sensitive to density. A measurement of the wind plasma density ($\log n=14$ cm$^{-3}$) has only been made for one source, BH LMXB GRO J1655$-$40, due to the presence of rare density-sensitive absorption features in its soft state spectrum \citep{miller08}. Simulations of accretion disks and thermally launched winds predict plasma densities at lower values in the range of $\log n=11-13$ cm$^{-3}$. Densities of $\log n=12$ cm$^{-3}$ are commonly used in disk wind studies, and we adopt this value in our analysis.

In XSTAR we simulated plasmas illuminated by different types of ionizing spectra across the 1eV to 100 MeV energy range, including a generic powerlaw ($\Gamma$ = 2), and our \emph{RXTE}/PCA$-$\emph{Chandra} HETGS GX 13+1 continuum. It is common to run XSTAR simulations with a powerlaw ionizing spectrum, and while LMXB spectra can be roughly approximated by a powerlaw, it is not a physically consistent model, especially across the broad energy range in which the ion populations are generated. A more realistic ionizing spectrum based on our continuum model included a multicolor disk, a blackbody plus a high-energy powerlaw component that is cut-off below the blackbody temperature.

\begin{figure}
\begin{tabular}{c}
   \includegraphics[scale=0.35, angle=270.]{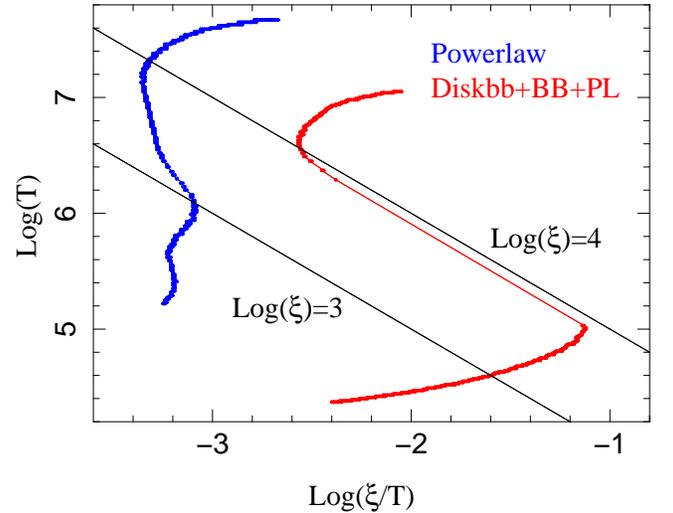} \\
\end{tabular}
   \caption{Thermal stability curves for two ionizing spectra. A powerlaw (blue curve) generates a plasma with a significantly higher temperatures than a thermal ionizing spectrum (red curve). The black lines are lines of constant ionization parameter ($\log \xi=3, 4$ for bottom/top, respectively); the thermal stability curves span ionization parameters $2\leq \log \xi \leq 5$, and both simulated plasmas have densities of $10^{12}$ cm$^{-3}$. \protect{\label{fig:thermstab}} }
\end{figure}

From the XSTAR output we generated thermal stability curves ($\log(\frac{\xi}{T})$ versus $\log(T)$) for different ionizing spectra, shown in Figure \ref{fig:thermstab}. Portions of the thermal stability curves with negative derivatives correspond to unstable solutions. It is generally agreed that one does not expect to observe ion signatures corresponding to plasmas with temperatures and ionization parameters along the unstable branch. This phenomenon has been invoked to explain the absence of ionized absorption features in BH LMXBs' hard state spectra. Unstable solutions correspond to a large range of unstable ionization parameters ($3.55 \leq \log \xi \leq 4.2$). This range happens to correspond with the peak fractions of many H-like and He-like ion species, making them essentially unobservable, which is in agreement with the absence of disk wind absorption features observed in the hard state \citep{chakravorty2013,bianchi17}.

From Figure \ref{fig:thermstab} it is obvious that  the nonthermal powerlaw ionizing spectrum produces very different plasma behavior (blue curve), occupying a temperature domain hotter for any given ionization parameter compared to a realistic thermal ionizing spectrum (red curve). As GX 13+1's spectrum did not change significantly between any of the observations, the ionizing spectrum input into XSTAR is essentially the same from one observation to another. To aid in comparing differences in \emph{warmabs} parameters between observations, we used the population file generated with the ObsID 11815 ionizing spectrum in all of our \emph{warmabs} fits.

\begin{table*}
   \centering
\begin{threeparttable}
   \caption{\textbf{Single \emph{warmabs} component fits} \label{tab:wasinglefit}}
   \centering
   \tabcolsep=0.1cm
   \scriptsize
\begin{tabular}{l c c c c c c c c c c r }
   \hline
   \hline
ObsID & $N_{\mathrm{H},22}$ & $\log \xi$ &  $v_{\mathrm{turb}}$ & $v_{\mathrm{out}}$ & Ca\tnote{a}& Fe & Ar & Si & S & Mg &  $\chi^2_{\nu}$ \ (dof)  \\
 & & & (km s$^{-1}$) & (km s$^{-1}$) & & & & & & & \\
   \hline
 & & & \multicolumn{6}{c}{$v_{\mathrm{turb}}$ Parameter Free} & & & \\
2708 & $25.9\pm1.1$ & $3.91\pm0.02$ & $200\pm20$ & $630\pm30$ & $1.77\pm0.28$ & $\hdots$ & $\hdots$ & $\hdots$ & $\hdots$ & $\hdots$ & $1.170 \ (6043)$ \\ 
11815 & $26.2\pm1.8$ & $3.92\pm0.01$ & $100\pm40$ & $650\pm70$ & $1.97\pm0.35$ & $\hdots$ & $\hdots$ & $\hdots$ & $\hdots$ & $\hdots$ & $1.148 \ (5966)$ \\ 
11816 & $25.6\pm3.0$ & $4.17\pm0.05$ & $210\pm10$ & $990\pm50$ & $1.98\pm0.79$ & $\hdots$ & $\hdots$ & $\hdots$ & $\hdots$ & $\hdots$ & $1.158 \ (5935)$ \\ 
11814 & $15.1\pm1.4$ & $3.90\pm0.04$ & $135\pm20$ & $540\pm50$ & $1.50\pm0.68$ & $\hdots$ & $\hdots$ & $\hdots$ & $\hdots$ & $\hdots$ & $1.145 \ (5827)$ \\ 
11817 & $24.8\pm1.8$ & $3.93\pm0.01$ & $130\pm30$ & $500\pm60$ & $1.27\pm0.41$ & $\hdots$ & $\hdots$ & $\hdots$ & $\hdots$ & $\hdots$ & $1.153 \ (5879)$ \\
   \hline
& & & \multicolumn{6}{c}{$v_{/mathrm{turb}}$ Parameter Fixed} & & & \\
2708 & $28.3\pm2.6$ & $4.04\pm0.06$ & $499$* & $740\pm40$ & $3.36\pm0.65$ & $\hdots$ & $\hdots$ & $\hdots$ & $\hdots$ & $\hdots$ & $1.214 \ (6044)$ \\ 
11815 & $15.6\pm1.1$ & $3.92\pm0.03$ & $490$* & $840\pm40$ & $3.23\pm0.61$ & $\hdots$ & $\hdots$ & $\hdots$ & $\hdots$ & $\hdots$ & $1.178 \ (5967)$ \\ 
11816 & $16.1\pm1.4$ & $4.15\pm0.08$ & $555$* & $870\pm10$ & $3.39\pm1.17$ & $\hdots$ & $\hdots$ & $\hdots$ & $\hdots$ & $\hdots$ & $1.165 \ (5936)$ \\ 
11814 & $14.1\pm1.1$ & $4.08\pm0.04$ & $481$* & $820\pm70$ & $4.12\pm1.16$ & $\hdots$ & $\hdots$ & $\hdots$ & $\hdots$ & $\hdots$ & $1.157 \ (5828)$ \\ 
11817 & $14.8\pm1.4$ & $3.96\pm0.01$ & $265$* & $570\pm50$ & $2.26\pm0.82$ & $\hdots$ & $\hdots$ & $\hdots$ & $\hdots$ & $\hdots$ & $1.181 \ (5880)$ \\
   \hline
& & & \multicolumn{6}{c}{Fe Abundance Free} & & & \\
2708 & $39.9\pm2.3$ & $3.96\pm0.02$ & $200\pm40$ & $620\pm30$ & $1.73\pm0.28$ & $0.64\pm0.11$ & $\hdots$ & $\hdots$ & $\hdots$ & $\hdots$ & $1.175 \ (6042)$ \\ 
11815 & $34.9\pm1.1$ & $3.95\pm0.03$ & $150\pm30$ & $740\pm10$ & $1.77\pm0.32$ & $0.58\pm0.26$ & $\hdots$ & $\hdots$ & $\hdots$ & $\hdots$ & $1.154 \ (5965)$ \\ 
11816 & $27.7\pm4.2$ & $4.14\pm0.06$ & $640\pm200$ & $1050\pm60$ & $1.91\pm1.00$ & $0.55\pm0.14$ & $\hdots$ & $\hdots$ & $\hdots$ & $\hdots$ & $1.160 \ (5934)$ \\ 
11814 & $28.4\pm1.6$ & $3.98\pm0.05$ & $420\pm80$ & $620\pm60$ & $1.38\pm0.43$ & $0.33\pm0.05$ & $\hdots$ & $\hdots$ & $\hdots$ & $\hdots$ & $1.142 \ (5826)$ \\ 
11817 & $32.0\pm1.8$ & $3.96\pm0.02$ & $150\pm10$ & $480\pm30$ & $1.07\pm0.33$ & $0.46\pm0.07$ & $\hdots$ & $\hdots$ & $\hdots$ & $\hdots$ & $1.152 \ (5878)$ \\
   \hline
& & & \multicolumn{6}{c}{Ar/S/Si/Mg Abundances Free} & & & \\
2708 & $24.7\pm1.3$ & $4.13\pm0.03$ & $400\pm40$ & $680\pm30$ & $4.99\pm0.88$ & $\hdots$ & $2.9\pm0.8$ & $4.4\pm0.5$ & $3.0\pm0.4$ & $3.2\pm1.0$ & $1.164 \ (6040)$ \\ 
11815 & $20.9\pm1.1$ & $4.10\pm0.03$ & $400\pm40$ & $780\pm40$ & $5.22\pm0.91$ & $\hdots$ & $3.0\pm0.9$ & $4.4\pm0.4$ & $3.0\pm0.4$ & $2.6\pm0.9$ & $1.148 \ (5962)$ \\ 
11816 & $18.0\pm1.6$ & $4.22\pm0.06$ & $680\pm160$ & $1030\pm60$ & $3.85\pm1.51$ & $\hdots$ & $3.8\pm1.5$ & $4.1\pm0.9$ & $2.2\pm0.8$ & $<1.3$ & $1.154 \ (5931)$ \\ 
11814 & $13.7\pm1.3$ & $4.14\pm0.05$ & $570\pm120$ & $630\pm60$ & $4.92\pm1.58$ & $\hdots$ & $1.5\pm1.8$ & $6.4\pm1.0$ & $3.2\pm0.8$ & $5.5\pm1.9$ & $1.134 \ (5824)$ \\ 
11817 & $21.7\pm1.5$ & $4.22\pm0.03$ & $210\pm10$ & $520\pm30$ & $3.71\pm1.21$ & $\hdots$ & $3.6\pm1.1$ & $5.9\pm0.7$ & $3.4\pm0.6$ & $3.2\pm1.3$ & $1.140 \ (5875)$ \\ 
   \hline
\end{tabular}
   \begin{tablenotes}
   \item[a] Calcium overabundance required in all \emph{warmabs} fits.
   \end{tablenotes}
\end{threeparttable}
\end{table*}

\subsection{Warm Absorber Fits}
\label{sec:waresults}

In our \emph{warmabs} modeling, we allowed for warm absorber column densities between $N_{\mathrm{H}}=10^{20-24}$ cm$^{-2}$, ionization parameters $\log\xi=2.5-5.0$, outflow and turbulent velocities in the range of $0-1500$ km s$^{-1}$. In fits with a single warm absorber, we were able to constrain the broad emission feature, but to reduce the number of variable parameters, we froze the broad emission parameters to the values found in Table \ref{tab:febroadparam}.

We found our \emph{warmabs} fits depended significantly on our incident ionizing spectrum. For a \emph{powerlaw} ionizing spectrum, one absorber could produce all seven accretion disk wind absorption features at their observed line widths and depths, while a more physically motivated ionizing spectrum of an accretion disk, a blackbody plus high-energy powerlaw, required multiple absorbers to produce all seven major features, unless we allowed for non-standard abundances.

\begin{figure}
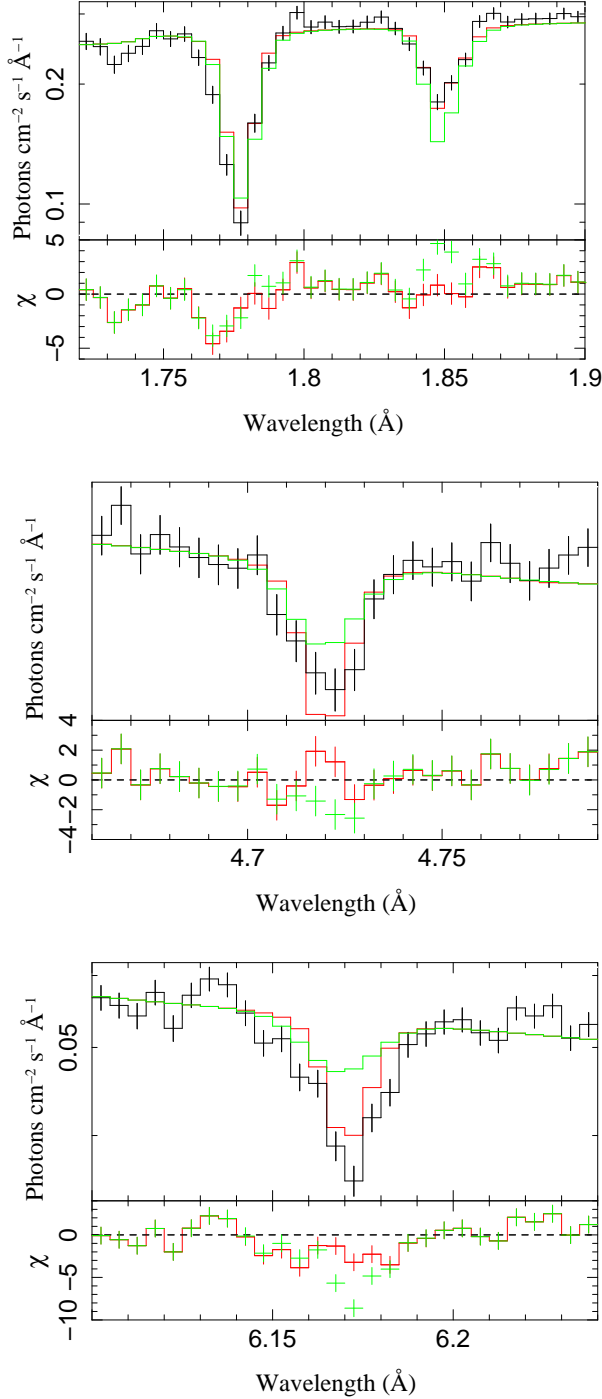

   \centering
\begin{tabular}{c}
   \includegraphics[scale=0.3, angle=270.]{11815_fe_region_paper.ps} \\
   \includegraphics[scale=0.3, angle=270.]{11815_si_region_paper.ps} \\
   \includegraphics[scale=0.3, angle=270.]{11815_s_region_paper.ps} \\
\end{tabular}
   \caption{\emph{Chandra} HETGS data for ObsID 11815 are shown in black for the iron line region (top), \ion{S}{16} (middle), and \ion{Si}{14} absorption features. The \emph{warmabs} model in red shows a fit with a single absorber and variable turbulent velocity; the iron lines are well fit, but the absorber has a small turbulent velocity ($v_{\mathrm{turb}}\approx120$ km s$^{-1}$) and the predicted sulfur and silicon lines are too narrow. In green, we plot the resulting fit when the \emph{warmabs} $v_{\mathrm{turb}}$ parameter is fixed to the turbulent velocity value calculated from the gaussian line fits ($490$ km s$^{-1}$); the produced sulfur and silicon absorption is far below what is observed. These results demonstrate that at least two absorbers are required to fit the iron, sulfur, and silicon lines to match their observed velocity-broadened widths and EWs, unless non-standard abundances are allowed. \protect{\label{fig:badfitexamp}} }
\end{figure}

With an ionizing spectrum based on extrapolating the $1.4-40$ keV \emph{Chandra} and \emph{RXTE} continuum model, a single \emph{warmabs} component can produce absorption close to the observed levels, but the correct \ion{Fe}{25}/\ion{Fe}{26} ratio is not produced, and the turbulent velocities (WA $v_{\mathrm{turb}}<200$ km s$^{-1}$) are a fraction of the values computed from our gaussian fits (FWHM$\implies$ $v_{\mathrm{turb}}\approx500$ km s$^{-1}$). Fit parameters for a single \emph{warmabs} component are shown in Table \ref{tab:wasinglefit}. The \emph{warmabs} fit with a variable turbulent velocity is shown in red in Figure \ref{fig:badfitexamp}; the lower ionization energy lines are visibly too narrow. We found this issue arose in all of the observations when fit with a single \emph{warmabs} component if standard abundances were assumed.

We calculated an average turbulent velocity for each observation based on our gaussian fits to the \ion{S}{16} and \ion{Si}{14} features as they are the strongest absorption lines in the portion of the spectrum with the highest spectral resolution. Fixing the \emph{warmabs} $v_{\mathrm{turb}}$ parameter to our gaussian velocity widths, the high-ionization parameters required to produce the iron lines ($\log\xi\approx$ 4.1), significantly under-fit the sulfur, silicon, and magnesium lines. Additionally, the $\chi^2_{\nu}$ values are significantly worse with the $v_{\mathrm{turb}}$ parameter fixed. In Figure \ref{fig:badfitexamp}, the \emph{warmabs} fit with a single absorber with a fixed turbulent velocity is shown in green.

All of the aforementioned \emph{warmabs} modeling have assumed standard Wilms solar abundances \citep{wilmsabund}, except for calcium, which is required to vary in all \emph{warmabs} fits. If we allow the warm absorber's abundances to vary (in addition to calcium's), one absorber can provide reasonable fits to all of the absorption features and match the expected turbulent broadening; details of the fits are listed in Table \ref{tab:wasinglefit}. Allowing only the iron abundance to vary requires iron abundances 30\% relative to solar, although the iron abundances are not consistent between observations and the \ion{Si}{14} absorption line is still underfit. With the iron abundance relative to solar fixed to unity and allowing the calcium, argon, sulphur, silicon, and magnesium (i.e. the other major absorption feature elements) abundances to vary, requires overabundances $(2-6)\times$ solar values, but provides a better fit than allowing only the iron abundance to vary. We acknowledge that a single absorber with super-solar Ca, Ar, S, Si, and Mg abundances is a possible solution to the absorption complex modeling, but we continued to search for solutions with fewer variable abundances, finding that multiple absorption zones were required.

\newpage

\begin{sidewaystable}
\begin{threeparttable}
   \caption{\textbf{Multiple \emph{warmabs} component fits} \label{tab:wasmultfit}}
   \centering
   \tabcolsep=0.07cm
   \scriptsize
\begin{tabular}{l c c c c c c c c c c c c c c c r }
   \hline
   \hline
ObsID & $N_{\mathrm{H},22}(1)$ & $\log \xi(1)$ &  $v_{\mathrm{turb}}(1)$ & $v_{\mathrm{out}}(1)$ & Mg\tnote{a} & $N_{\mathrm{H},22}(2)$ & $\log \xi(2)$ & $v_{\mathrm{turb}}(2)$ & $v_{\mathrm{out}}(2)$ & Ca\tnote{a} & $N_{\mathrm{H},22}(3)$ & $\log \xi(3)$ &  $v_{\mathrm{turb}}(3)$  &  $v_{\mathrm{out}}(3)$ & Ca\tnote{a} & $\chi^2_{\nu}$ \ (dof)  \\
  & &  & (km s$^{-1}$) & (km s$^{-1}$) & & & & (km s$^{-1}$) & (km s$^{-1}$) & & & & (km s$^{-1}$) & (km s$^{-1}$) & & \\
   \hline
 & & & & \multicolumn{5}{c}{Two \emph{warmabs} Components} & & & \\
2708 & $0.8\pm0.1$ & $2.98\pm0.03$ & $290\pm80$ & $570\pm40$ & $0.52\pm0.22$ & $22.1\pm1.8$ & $4.10\pm0.04$ & $600\pm140$ & $860\pm60$ & $5.35\pm1.11$ & $\hdots$ & $\hdots$ & $\hdots$ & $\hdots$ & $\hdots$ & $1.143 \ (6032)$ \\ 
11815 & $0.6\pm0.1$ & $2.92\pm0.01$ & $380\pm110$ & $630\pm60$ & $0.34\pm0.19$ & $17.7\pm0.7$ & $3.99\pm0.02$ & $410\pm20$ & $910\pm50$ & $4.07\pm0.70$ & $\hdots$ & $\hdots$ & $\hdots$ & $\hdots$ & $\hdots$ & $1.132 \ (5962)$ \\ 
11816 & $0.3\pm0.1$ & $2.86\pm0.10$ & $250\pm210$ & $790\pm110$ & $0.28\pm0.25$ & $17.8\pm1.5$ & $4.21\pm0.05$ & $820\pm190$ & $1170\pm80$ & $3.80\pm1.71$ & $\hdots$ & $\hdots$ & $\hdots$ & $\hdots$ & $\hdots$ & $1.141 \ (5924)$ \\ 
11814 & $0.5\pm0.1$ & $2.71\pm0.05$ & $260\pm110$ & $310\pm50$ & $0.59\pm0.22$ & $12.5\pm1.3$ & $4.07\pm0.05$ & $380\pm90$ & $960\pm80$ & $4.48\pm1.63$ & $\hdots$ & $\hdots$ & $\hdots$ & $\hdots$ & $\hdots$ & $1.111 \ (5816)$ \\ 
11817 & $0.7\pm0.1$ & $2.88\pm0.07$ & $60\pm130$ & $400\pm60$ & $0.31\pm0.21$ & $13.5\pm1.3$ & $4.05\pm0.05$ & $400\pm150$ & $760\pm70$ & $3.32\pm1.20$ & $\hdots$ & $\hdots$ & $\hdots$ & $\hdots$ & $\hdots$ & $1.120 \ (5870)$ \\
   \hline
& & & & \multicolumn{5}{c}{Three \emph{warmabs} Components\tnote{b}} & & & \\
2708 & $0.8\pm0.1$ & $2.94\pm0.1$ & $340\pm510$ & $590\pm320$ & $0.44\pm0.3$ & $0.2\pm0.1$ & $3.57\pm0.11$ & $340$\tnote{c} & $590$\tnote{c} & $5.60\pm1.65$& $30.4\pm12.1$ & $4.25\pm0.10$ & $630\pm370$ & $890\pm60$ & $5.60$\tnote{d} & $1.146 \ (6031)$ \\ 
11815 & $0.7$ & $3.01$ & $390$ & $640$ & $0.35$ & $0.1$ & $3.50$ & $390$ & $640$ & $4.55$ & $32.2$ & $4.21$ & $420$ & $960$ & $4.55$ & $1.139 \ (5961)$ \\ 
11816 & $0.4$ & $2.93$ & $420$ & $790$ & $<0.50$ & $0.2$ & $3.39$ & $420$ & $790$ & $3.60$ & $19.2$ & $4.25$ & $850$ & $1190$ & $3.60$ & $1.149 \ (5923)$ \\ 
11814 & $0.7$ & $3.03$ & $390$ & $390$ & $0.65$ & $0.1$ & $3.45$ & $390$ & $390$ & $6.11$ & $20.1$ & $4.32$ & $630$ & $1030$ & $6.11$ & $1.116 \ (5815)$ \\
11817 & $0.7$ & $3.03$ & $130$ & $480$ & $0.33$ & $0.2$ & $3.46$ & $130$ & $480$ & $3.69$ & $22.1$ & $4.31$ & $420$ & $840$ & $3.69$ & $1.136 \ (5869)$ \\ 
   \hline
\end{tabular}
   \begin{tablenotes}
   \item[a] Variable calcium and magnesium abundances are required.
   \item[b] The 90\% error bounds for the three warmabs fits were estimated with Markov chain Monte Carlo runs. Bounds for the other observations are similar to those stated for ObsID 2708.
   \item[c] The turbulent and blueshift velocities were tied between the WA(1) and WA(2) components for the 3 \emph{warmabs} components model.
   \item[d] The WA(2) and WA(3) calcium abundances were tied.
   \end{tablenotes}
\end{threeparttable}
\end{sidewaystable}

The \ion{Fe}{26} absorption feature's large EW ($30-45$ eV) and, in particular, its width relative to \ion{Fe}{25}, was the first indication that a warm absorber with ionization parameter $\log \xi \gtrsim$4 is present in GX 13+1's spectrum. All accretion disk wind LMXBs show absorption lines from \ion{Fe}{25} and/or \ion{Fe}{26}, which can be modeled by photoionized plasmas with $\log\xi\gtrsim$3. Unlike most BH and NS disk wind sources where \emph{only} iron lines are observed, GX 13+1's spectrum exhibits significant absorption from ions with lower ionization parameters (for example, the \ion{Si}{14} K$\alpha$ absorption line is at $\approx2$ keV with EWs of $1.2-2.9$ eV), indicating a less ionized plasma is also present.

Adding more \emph{warmabs} components, we found the narrow absorption features could be successfully modeled with two or three warm absorbers, see Table \ref{tab:wasmultfit}. In both scenarios, the calcium and magnesium abundances were allowed to vary in order for their full EWs to be modeled. Fitting with two \emph{warmabs} components, the iron lines are associated with a highly ionized absorber ($\log \xi \approx 4.1$, $N_{\mathrm{H},22}\approx10-20$), while the silicon, sulfur, magnesium, calcium, and argon lines are partially produced by a much less ionized absorber ($\log \xi\approx$2.9, $N_{\mathrm{H},22}\approx0.3-0.8$). The two components exhibit similar turbulent velocities ($v_{\mathrm{turb}}\approx100-800$ km s$^{-1}$), while the more highly ionized absorber has a larger blueshift ($v_{\mathrm{outflow}}\approx750-1150$ km $s^{-1}$) than the less ionized component ($v_{\mathrm{outflow}}\approx300-800$ km s$^{-1}$).

In a model with three \emph{warmabs} components, which may not be a unique solution, one highly ionized absorber ($\log\xi\approx4.25$, $N_{\mathrm{H,22}}\approx20-30$) generates almost exclusively \ion{Fe}{26} K$\alpha$ absorption, a second absorber ($\log\xi\approx3.5$, $N_{\mathrm{H,22}}\approx0.1-0.4$) produces significant \ion{Fe}{25} absorption along with some of the lower energy lines (\ion{S}{16}, \ion{Si}{14}, etc.) and a third absorber produces ($\log\xi\approx2.95$, $N_{\mathrm{H,22}}\approx0.4-0.8$) the rest of the \ion{S}{16}, \ion{Si}{14}, and \ion{Mg}{12} absorption features. Similar to our findings with a two \emph{warmabs}-component-model, the component associated with the \ion{Fe}{26} line had a larger blueshift ($v_{\mathrm{out}}=800-1200$ km s$^{-1}$) than the two components associated with the lower ionization lines ($v_{out}=350-800$ km s$^{-1}$). For both the two- and three-\emph{warmabs}-component solutions, we find the lowest ionization component ($\log\xi \approx2.9$) produces \ion{Si}{13} absorption around 1.865 keV (6.65 \AA) in the silicon edge, as predicted by \citet{schulzsik}.

For the three \emph{warmabs} fits, we assessed the error bars by performing a Markov chain Monte Carlo (MCMC) analysis using a code implemented for the \texttt{ISIS} package based upon the methods described in \citealt{foreman}.  (See the description of this \texttt{ISIS} script in \citealt{murphy}.)  We evolved a set of 320 ''walkers'' (10 per free parameter in the fit) until the MCMC probability distributions reached equilibrium, as judged by the parameter probability histograms from the final quarter of steps in the chain being nearly identical to the probability distributions from the third quarter of chain steps;  90\% confidence level error bars were then calculated from the parameter probability distributions using the final third of chain steps.

\subsection{Variability and the RXTE CD}
\label{sec:reswindvar}

Accretion disk wind absorption line variability, attributed to changes in accretion flow, has been seen on both short ($\approx$ ks) and long (days$\rightarrow$months) timescales. On timescales down to tens of seconds, \citet{neilsen11} correlated dramatic cyclical changes with a period of $\approx50$ s in the illuminating continuum to changes in the accretion disk wind column density in GRS 1915+105. On timescales of tens of ks in GRS 1915+105, \citet{lee01} found changes in the accretion disk flux and the accretion disk wind density drive changes in the iron absorption features.

In previous analyses of GX 13+1, variability has been claimed on day and kilosecond timescales \citep{ueda2004,dtrigoxmm}. While \citet{sidoli2002} found no significant changes on day timescales in the \ion{Fe}{25} and \ion{Fe}{26} K$\alpha$ EWs, \citet{dtrigoxmm} found significant correlations between the hard flux ($6-10$ keV) and the ionization state and column density of the absorber in their five \emph{XMM}/EPIC observations, which were taken days to weeks apart from one another. \citet{ueda2004} claimed variability in the absorber's ionization on timescales of five ks in the 2002 HETGS observation of GX 13+1; changes in the iron lines' ratio appeared to be correlated with the continuum flux.

We compared our measured wind properties with GX 13+1's position along its Z tracks, which are plotted in Figure \ref{fig:ztrack}. All of the \emph{Chandra} observations occurred while GX 13+1 was on the NB and HB; blueshifted absorption features detected in all of these observations suggests these branches are both strongly associated with an accretion disk wind. Strong radio emission commonly associated with jets has also been observed at the vertex of the NB and HB along one of the Z tracks \citep{homan04}, which suggests accretion disk winds and jets can be simultaneous in LMXBs and is the subject of the work of \citet{homan16}.

During several of the \emph{Chandra} observations, GX 13+1 exhibited significant movement along the CD. While the GX 13+1 light curves do not suggest significant changes or trends in the source's intensity during the observations despite the known motion along the Z track, we investigated whether the accretion disk wind exhibits variability on small timescales as reported by \citet{ueda2004} by breaking the observations into 10 ks time segments, the minimum amount time required to achieve meaningful constraints on the flux and EWs.

\begin{figure}[h!]
\begin{tabular}{c}
   \includegraphics[scale=0.35,angle=270]{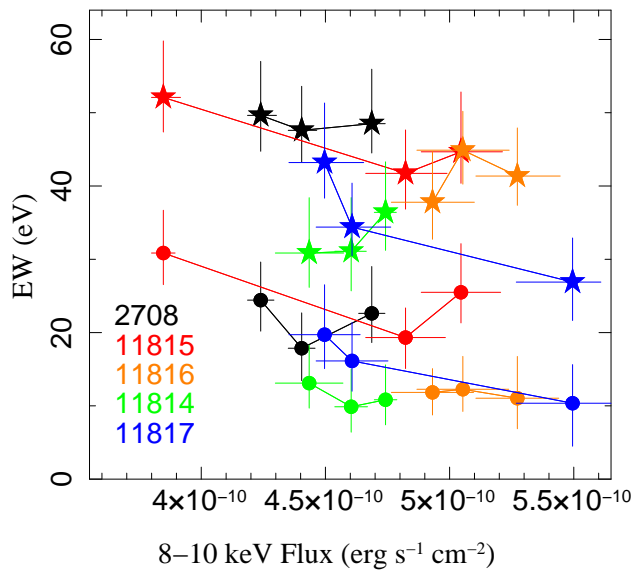} \\
\end{tabular}
   \caption{ \ion{Fe}{26} (star symbols) and \ion{Fe}{25} (filled circles) EWs versus the $8-10$ keV flux in 10 ks time steps of each of the five TE mode observations. The change in hard flux during the course of the observation roughly reflects the source's motion along its Z track. In ObsID 11816, GX 13+1 occupies a relatively small area at the HB/NB vertex, which is reflected in the small range of its $8-10$ keV flux across the time segments.\protect{\label{fig:timechunk10}}}
\end{figure}

For each time segment, we fit the \emph{Chandra} HETGS data with a continuum model, \emph{simple\_gpile2 [ tbnew $\times$ (warmabs $\times$ (diskbb+bbodyrad+powerlaw))+emission-absorption]}, where the emission is the broad gaussian emission frozen to the values in Table \ref{tab:febroadparam}, and the absorption is the \ion{Fe}{26} and \ion{Fe}{25} narrow gaussian features. We plotted the powerlaw normalization as a function of the \emph{RXTE} orbit and did not find significant variability within any observation allowing us to fix the powerlaw normalization to the values in Table \ref{tab:contparam}. We calculated the \ion{Fe}{26} and \ion{Fe}{25} K$\alpha$ lines' EWs and plotted them against the hard $8-10$ keV flux in each 10 ks segment (see Figure \ref{fig:timechunk10}).

Overall, both the \ion{Fe}{26} and \ion{Fe}{25} EWs exhibit a correlation with the $8-10$ keV flux; as the hard flux increases, the EWs also decrease. Looking at individual observations, the hard flux evolves across the 10 ks segments as expected with their movement along the CD (see Figure \ref{fig:ztrack}). For example, in ObsID 11817 GX 13+1 moves along its Z track from the NB/HB vertex to the NB/FB vertex, during which both the soft and hard colors decrease. In Figure \ref{fig:timechunk10}, GX 13+1 can be found in the first 10 ks of the observation with an $8-10$ keV flux of $5.2\times10^{10}$ erg s$^{-1}$ cm$^{-2}$; the hard flux decreases throughout the observation as the  \ion{Fe}{26} and \ion{Fe}{25} EWs both increase. The \ion{Fe}{26} EW increases by over 50\% while the \ion{Fe}{25} EW nearly doubles.

In a disk wind model where two absorbers are present, one absorber produces both of the Fe K$\alpha$ lines. The EW ratio (\ion{Fe}{25} EW / \ion{Fe}{26} EW) is a probe of the ionization state. The errors in the EW ratios, however, are too large to interpret relative changes in the absorption features during any of the observations. The apparent correlation between both the iron lines' EW and the $8-10$ keV flux does not support a change in the ionization state. For both the \ion{Fe}{26} and on \ion{Fe}{25} EWs to decrease with the hard flux, the ionization state must remain remain constant while the wind column density decreases. If one absorber produces the \ion{Fe}{26} line and another produces the \ion{Fe}{25} line, as is the case in our three-absorber model, the trend could be produced by several scenarios involving changes in both the ionization states and/or the column densities.

\section{Discussion}
\label{sec:discussion}
As the number of high-resolution observations of disk wind systems has grown over the past decade, it has become increasingly clear that disk winds play a critical role in the overall accretion process in both BH and NS LMXBs. GX 13+1 is one of only three NS narrow absorption line systems that shows a definitive outflow; IGR J17480-2446 \citep{millerigrj17480} and Cir X-1 \citep{brandt2000} are the others. In contrast, all BH narrow line systems exhibit blueshifts \citep{trigoboirin}. GX13+1 is bright, accreting at more than 50\% Eddington and displays seven major absorption features, the most amongst NS disk wind systems, revealing a detailed view of the absorbing plasma's properties. Its simultaneous \emph{Chandra} HETGS and \emph{RXTE}/PCA observations offer the unique chance to study the disk wind properties along the source's horizontal and normal branches.

The abundance of high-S/N observations of GX 13+1 have allowed us to perform a detailed analysis of the disk wind absorption spectrum in GX 13+1. We have fit the \emph{Chandra} and \emph{RXTE} $1.5-40$ keV spectrum and performed direct line fits of the seven strongest absorption features, including hydrogen- and helium-like iron K$\alpha$ lines and the K$\alpha$ lines from hydrogen-like Ca, Ar, S, Si, and Mg. Through careful photoionization modeling of the absorption features, we found multiple absorbers were required to produce the observed disk wind signature, unless nonstandard abundances were assumed.

\subsection{Multiple Absorbers in Accretion Disk Winds}

Previous analyses of the accretion disk wind in GX 13+1 required only one warm absorber, while we find at least two absorbers are necessary. The disagreement may be a result of inherent variability or our improved treatment of the ionizing spectrum. In the five \emph{XMM} observations \citet{dtrigoxmm} found significant absorption due to the K$\alpha$ and K$\beta$ transitions of \ion{Fe}{25} and \ion{Fe}{26}. Evidence of absorption due to the lower energy transitions in \ion{Si}{14}, \ion{Si}{13}, \ion{S}{15}, \ion{S}{16}, and \ion{Ca}{20} were present in several observations. The measured EWs were compatible with a single, highly ionized absorber with $\log\xi=4$ \citep{dtrigoxmm}. If only one absorber was indeed present in the \emph{XMM} observations, it would indicate variability in the wind's ionization state, as the lower-ionization components we see in our \emph{Chandra} observations are absent. However, the low energy resolution of the \emph{XMM}/EPIC observations makes the detection of the narrow, less ionized absorption features unlikely even if they were present.

In the best fits to our data, we found at least two absorption zones with different ionization parameters and outflow velocities were required to produce the observed disk wind signature. Multiple absorbers have been seen in disk wind systems before. Evidence of multiple velocity components was seen in LMXB BH GRO J1655-40 \citep{kallman09,neilsen12}. Taking a closer look at the iron line region ($6-8$ keV) in the \emph{third-order} HEG spectra of four BH disk wind systems, including GRO J1655-40, \citet{miller15multabs} found fits with two or more absorption zones, significantly improved their fits. 

Specifically, among NS systems with narrow line absorption features, multiple absorption zones are not uncommon; three have one low-ionization component ($\log\xi < 3$) and one high-ionization component ($\log\xi > 3$) \citep{trigoboirin}. Absorption features associated with ions observed in GX 13+1's spectrum, including \ion{S}{16}, \ion{S}{14}, and \ion{Mg}{12}, have also been observed in the NS binaries 4U 1916-05 and Cir X-1. In 4U 1916-05, an accretion disk atmosphere source, the \ion{Fe}{25} and \ion{Fe}{26} features were produced by an absorber with $\log\xi \approx 4.15$ while the S, Si, Mg, and Ne lines were associated with a much less ionized component, $\log\xi \approx 3$ \citep{iaria06xb1916}. Similarly, in GX 13+1 we found absorption zones with $\log\xi=2.9-3.5$ produced the less ionized features. While the absorption features in Cir X-1 were very broad (FWHM $\approx$ 2000 km s$^{-1}$ compared to the 400-600 km s$^{-1}$ we observe in GX 13+1), they exhibited strong P-Cygni profiles, supporting the interpretation that the lines are due to an accretion disk wind \citep{brandt2000,schulz2002}. 

Observing multiple absorption zones in an accretion disk wind will provide multiple measurements of the plasma's outflow velocities and ionization parameters and will help probe the disk wind structure and, hence, the wind launching mechanism. Our analysis highlights the importance of not using an approximated ionizing spectrum, such as a powerlaw, in the XSTAR analysis. A re-analysis of the warm absorber modeling in disk wind sources may reveal multiple absorption components previously not identified. Future X-ray instruments with higher energy resolution and larger effective area will allow us to see more and more absorption line components, as weak lines become visible and blended lines become resolved \citep{kallman09}. 

In disk wind and disk atmosphere systems where multiple absorbers provide the best fit the to absorption line complex, the absorption zones are usually typified by distinct ionization states or by outflow velocity. In GX 13+1, the absorption zones differ by the associated column, ionization state, and velocity. These multiple absorption zones may be a natural signature of a smooth outflow. Simulations of Compton-heated winds suggest the absorption line complex may be characterized by two or three ionized zones \citep{giustini12,higproga15}. Alternatively, the different ionization zones may correspond to "clumps" of outflowing material, challenging the assumptions of a continuous and homogenous outflow. In this scenario, one might expect strong temporal variability in the absorption line features due to the noncontinuous outflow of absorbing material. As we see the same absorption features in all seven observations that span almost 10 years of monitoring, and model the absorption zones with similar ionization parameters and column densities, this suggests a relatively continuous and smooth outflow as opposed to large, separate physical clumps of outflowing material.

\subsection{Disk Wind Launching Mechanism}

Compton heating and magnetic driving are considered the most viable mechanisms in X-ray binary disk wind systems. For several BH wind systems (e.g. GRO J1655-40 and GRS 1915+105), there have been claims of magnetic driving based on small inferred wind launching radii (\citealt{miller06,miller16}, but see also \citealt{neilsencomptonthick} and \citealt{shidatsu} for an alternative explanation). Unusual wind properties or changes in wind properties within a single outburst have also driven the formulation of hybrid wind theories \citep{neilsen12}.

Previous studies of the disk wind in GX 13+1 have found that it is consistent with Compton heating \citep{sidoli2002,dtrigoxmm,dtrigoxmm,madej14}, although \citet{ueda2004} proposed radiation-driven wind based on a model where radiation driving could be effective at sub-Eddington luminosities. The question of the wind launching mechanism in any disk wind system is often answered by determining the wind launching radius ($R_{L}= \sqrt{\frac{L}{n \xi}}$); this is approximated as the radius where we observe the innermost absorption region which is estimated from the definition of the ionization parameter (Eqn. \ref{eqn:ionparam}). As previously discussed in Section \ref{sec:inputspec}, the largest uncertainty in disk wind observations is the plasma density. It is also the largest uncertainty in determining the wind launching radius and, hence, the wind mechanism. We have no independent measurement of the plasma density in GX 13+1, so for this reason and for consistency we use the same electron density we used to calculate our population levels in XSTAR, $n=10^{12}$ cm$^{-3}$.

Uncertainty in the bolometric luminosity is the second largest source of error in determining the wind launching radius. The bolometric luminosity is estimated from the X-ray luminosity, which can be significantly underestimated when there is a high opacity along the line of sight, the exact conditions when observing highly inclined disk wind systems \citep{trigoboirin}. While there is a large neutral column density ($N_{\mathrm{H},22}>1$) toward GX 13+1, as well as a warm absorber with an even larger column density ($N_{\mathrm{H},22}\approx30$), we benefit from simultaneous \emph{RXTE} observations in estimating a broadband flux. \citet{homan16} found that GX 13+1 was accreting at nearly its Eddington limit, $L=(1.2-1.3)\times10^{38}$ erg s$^{-1}$ ($\approx0.7-0.8$ $L_{\mathrm{Edd}}$) across 0.1-100 keV, while in this work we estimated slightly lower Eddington fractions ($\approx0.45$ $L_{\mathrm{Edd}}$), in part because we did not include the powerlaw flux below $\sim2.5$ keV. Both estimates of the bolometric luminosity are likely a \emph{lower} bound because the accretion disk's flux has not been corrected for inclination and because of the luminosity suppression associated with the source's high obscuration, as previously mentioned. The highest ionization parameter we detected in our observations of GX 13+1 was $\log \xi \approx4.3$. Rewriting the launching radius in terms of our estimates for the density and luminosity, we find

\begin{equation*}
   \label{eqn:windlaunchscale}
   \begin{multlined}
   R_{\mathrm{L}} = 7 \times 10^{10} \times \left(\frac{L}{1.3\times10^{38} \ \mathrm{erg \ s^{-1}}} \right)^{\frac{1}{2}} \\ \times \left(\frac{n}{10^{12} \ \mathrm{cm^{-3}} }\right)^{-\frac{1}{2}} \times \left(\frac{\xi}{10^{4.3} \ \mathrm{erg \ cm \ s^{-1}}}\right)^{-\frac{1}{2}} \mathrm{cm}.
   \end{multlined}
\end{equation*}

Our launching radius is consistent with other published estimates, $R_{\mathrm{L}}\approx10^{10-11}$ cm \citep{ueda2004,dtrigoxmm,dai14dip}, despite the variety of luminosities and densities used to estimate the radius. Compton-heated winds can be launched outside of the 0.1$\times$ Compton radius, $R_{\mathrm{C}}$ \citep{begel83}. The Compton radius, $R_{\mathrm{C}}$, is given by:

\begin{equation}
   \label{eqn:comprad}
   R_C=10^{10} \times \left( \frac{M_{NS}}{M_{\bigodot}} \right) \times \left(\frac{T_C}{10^8 \ \mathrm{K}} \right)^{-1} \ \mathrm{cm},
\end{equation}

\noindent
where $M_{\mathrm{NS}}$ is the NS mass and $T_{\mathrm{C}}$ is the Compton temperature defined as:

\begin{equation}
   \label{eqn:comptemp}
   T_{C}=\frac{1}{4k_B} \frac{\int_0^{\infty} h\nu L_{\nu} d\nu}{\int_0^{\infty} L_{\nu} d\nu}.
\end{equation}

From our \emph{Chandra} observations, we calculate a typical Compton temperature of $1.3\times10^7$ K (see the red curve in Figure \ref{fig:thermstab}), which translates to a Compton radius of $8\times10^{10}$ cm and is comparable with our wind launching radius. As a Compton-driven wind can be launched just beyond 0.1 $R_{\mathrm{C}}$, this means a Compton wind can be launched an order of magnitude closer to the central NS than our calculated wind launching radius, $R_{\mathrm{L}}$.

As \citet{ryota18} has recently demonstrated in the particular case of GX 13+1, at high Eddington fractions, electron (radiation) pressure likely contributes to the wind driving. The electron pressure reduces the local effective gravity and, hence, the local escape velocity; this effect makes it easier for Compton-heated winds to be launched at smaller and smaller radii \citep{progakallman02,homan16,done16,ryota18}. Applying an approximate theoretical correction factor to the Compton radius due to the effects of radiation pressure (see Eqn. 10 of \citealt{done16}), the Compton radius moves inwards as the source brightens while the launching radius moves outwards ($R_{L} \propto L^{\frac{1}{2}}$). As GX 13+1 has a luminosity of at least 0.5 $L_{\mathrm{Edd}}$, it is well within the regime where electron pressure becomes significant \citep{ryota18}. The Compton radius is likely at least an one order of magnitude smaller than our estimate with no electron pressure taken into consideration, making it even more robust to the Compton-heated wind criterion $R_{\mathrm{L}}> R_{\mathrm{C}}$. Additionally, \citet{ryota18} demonstrated that a simplistic Compton-radiation driven wind (i.e. a wind with a hybrid driving mechanism) can begin to produce the iron absorption and emission line features in the $\approx6.5-7$ keV spectrum.

The estimated wind launching radius is also well within the accretion disk extent, $R_{\mathrm{C}} < R_{\mathrm{L}} < R_{/mathrm{Disk}}$. The disk size can be estimated from the radius of the Roche Lobe ($R_{\mathrm{Lb}}$) as $R_{\mathrm{Disk}}\approx0.8 R_{\mathrm{Lb}}$. For an orbital period of 24 days, a NS mass of 1.4 M$_{\odot}$ and a binary mass ratio, $q$, of 1 lead to a semimajor axis, $a$, of $3.5\times10^{12}$ cm. We estimate a Roche lobe radius of approximately $1.3\times10^{12}$ cm \citep{rochelobe} and, hence, an approximate accretion disk extent of $10^{12}$ cm, which is at least an order of magnitude larger than our wind launching radius. Thus, the condition $R_{\mathrm{C}} < R_{\mathrm{L}} < R_{\mathrm{Disk}}$ is met and the accretion disk wind in GX 13+1 is consistent with a Compton-heated wind.

\subsection{Wind Mass Loss}

GX 13+1 is one of the few NS systems with an accretion disk wind and the outflow is consistent with a Compton wind. In general, if the accretion disk is too small relative to the Compton radius, a Compton-heated wind cannot be launched although mass loss via jets is still possible. The accretion disk size may explain why blueshifts are not detected in $\approx 70\%$ of high-inclination NS LMXBs that exhibit narrow absorption features \citep{trigoboirin}, indicating an ionized plasma is present but is in a static configuration as an accretion disk atmosphere rather than an outflow. Indeed, comparing the Compton radii and the accretion disk size in seven BH and NS warm absorber systems, including GX 13+1, \citet{trigoboirin} found systems whose absorption lines are not measurably blueshifted have disk sizes comparable or significantly less than the Compton wind launching radius ($R_{\mathrm{Disk}}\lesssim 0.1 R_{\mathrm{C}}$).

Constraining the fraction of NS LMXBs that have disk outflows is critical to understanding the contribution of disk winds to galactic feedback. The mass driven from the binary system by a radial wind is given by 

\begin{equation}
   \label{eqn:windmass}
   \dot{M}_{Wind} = 4 \pi m_p v_{outflow} \frac{L}{\xi} \frac{\Omega}{4\pi}.
\end{equation}

\noindent Assuming a maximal covering fraction of unity will give us an upper limit on the mass loss rate. As before, we will use the plasma properties for the most ionized component in GX 13+1 and a bolometric luminosity of $1.2\times10^{38}$ erg s$^{-1}$ \citep{homan16} with a v$_{\mathrm{outflow}}$ of 1000 km s$^{-1}$, we find $\dot{M}_{Wind}  < 2\times10^{-7}M_{\odot}$ yr$^{-1}$. For a more realistic disk wind covering fraction of $\frac{\Omega}{4\pi}\approx 0.4$, we estimate $\dot{M}_{Wind} \approx 6\times10^{-8}$ $M_{\odot}$ yr$^{-1}$. This estimate assumes a homogenous and symmetric absorber. Other studies of GX 13+1 have inferred disk wind outflow rates between $(0.8-2)\times10^{-8}$ $M_{\odot}$ yr$^{-1}$ \citep{ueda2004,dtrigoxmm}, but still comparable to the mass accretion rate. 

The wind's kinetic luminosity is given by 

\begin{equation}
   L_{kinetic} = \frac{1}{2} \dot{M}_{Wind} \ v^2_{outflow}
\end{equation}

\noindent
so even though the wind carries substantial mass from the system, the relatively low outflow velocity, using 1000 km s$^{-1}$ as an upper limit, translates to a kinetic luminosity of $L_{\mathrm{kinetic}} \approx 10^{-4}$ $L_{\mathrm{Edd}}$. This finding has been seen in the other BH and NS disk wind systems \citep{ponti16}. While the low kinetic luminosity may mean accretion disk winds do not significantly contribute to galactic feedback, the kinetic luminosity of the wind relative to the source luminosity is consistent with a Compton-heated wind. The kinetic luminosity of a Compton wind cannot exceed the luminosity intercepted by the outer disk, which heats the accretion disk and powers the wind. Assuming a disk flaring angle of $10^{\circ}$ and efficiencies of $1-10$\%, converting the illuminating luminosity to kinetic power predicts Compton disk wind kinetic luminosities in the range of $10^{-5.5}\rightarrow10^{-3.5}$ $L_{\mathrm{Edd}}$ (see Figure 3 in \citealt{ponti16}).

\subsection{Future Investigations}

GX 13+1 is a complex system and worthy of further study. There is increasing evidence that plasma thermal instabilities could explain the lack of disk wind signatures in the hard state \citep{chakravorty2013,bianchi17}. While GX 13+1's plasma heating curve exhibits a thermal instability, it does not exclude a large range of ionization parameters, and an accretion disk wind is still active even during GX 13+1's hardest spectral states. As the instability regions depend strongly on the ionizing spectrum, an in-depth analysis should be performed for a wider variety of NS outburst spectra. 

Finally, GX 13+1 is expected to exhibit an accretion disk wind while on the flaring branch of its Z track. High-resolution, soft X-ray observations and joint hard X-ray coverage would allow for the detection of the disk wind absorption features and determine the CD position. These observations could produce a picture of the softer state accretion disk wind, whose properties could then be compared to this work's analysis of the accretion disk wind while the source was on the spectrally hard horizontal and normal branches.

\section{Acknowledgements}

This research has made use of data obtained from the Chandra Data Archive and the Chandra Source Catalog, and software provided by the Chandra X-ray Center (CXC) in the application packages CIAO, ChIPS, and Sherpa. We would also like to thank Tim Kallman for his assistance with XSTAR.

\bibliography{references}

\end{document}